\documentclass[
 reprint,
 amsmath,amssymb,
 aps,
 prl
]{revtex4-1}

\usepackage{graphicx}
\usepackage{dcolumn}
\usepackage{bm}
\usepackage{color}
\usepackage{mathrsfs}
\usepackage{hyperref}

\begin{document}

\preprint{APS/123-QED}

\title{Quantum Metrology with Higher-order Exceptional Points in Atom-cavity Magnonics}
\author{Minwei Shi$^{1}$}

\author{Guzhi Bao$^{1,2}$}
\email{guzhibao@sjtu.edu.cn}

\author{Jinxian Guo$^{1,2}$}
\email{jxguo@sjtu.edu.cn}

\author{Weiping Zhang$^{1,2,3,4}$}
\email{wpz@sjtu.edu.cn}

\affiliation{$^{1}$ Tsung-Dao Lee Institute, and School of Physics and Astronomy,  Shanghai Jiao Tong University, Shanghai, 200240, China}

\affiliation{$^{2}$ Shanghai Research Center for Quantum Sciences, Shanghai, 201315, China}

\affiliation{$^{3}$ Shanghai Branch, Hefei National Laboratory, Shanghai, 201315, China}

\affiliation{$^{4}$ Collaborative Innovation Center of Extreme Optics, Shanxi University, Taiyuan, 030006, China}

\begin{abstract}
Exceptional points (EPs), early arising from non-Hermitian physics, significantly amplify the system's response to minor perturbations, 
and act as a useful concept to enhance measurement in metrology. In particular, such a metrological enhancement grows dramatically with the EP's order. 
However, the Langevin noises intrinsically existing in the non-Hermitian systems diminish this enhancement.
In this study, we propose a protocol for quantum metrology with the construction of higher-order EPs (HOEPs) in atom-cavity system through Hermitian magnon-photon interaction.
The construction of HOEPs utilizes the atom-cavity non-Hermitian-like dynamical behavior but avoids the external Langevin noises via the Hermitian interaction.
A general analysis is exhibited for the construction of arbitrary $n$-th order EP (EPn).
As a demonstration of the superiority of these HOEPs in quantum metrology, we work out an EP3/4-based atomic sensor with sensitivity being orders of magnitude higher than that achievable in an EP2-based one.
We further unveil the mechanism behind the sensitivity enhancement from HOEPs.
The experimental establishment for this proposal is suggested with potential candidates.
This EP-based atomic sensor, taking advantage of the atom-light interface, offers new insight into quantum metrology with HOEPs.
\end{abstract}

\maketitle

\textit{Introduction.}-- 
In the past decades, new approaches for precision measurement have been proposed based on exceptional points (EPs), where the $n$-th order EP (EPn) indicates the coalescence of the system’s $n$ ($n\geq 2$) eigenvalues and their corresponding eigenstates \cite{miri2019exceptional,heiss2012physics,ozdemir2019parity,peng2016anti,liang2023observation}. 
The system exhibits a nonlinear response to external perturbations $\epsilon$ near EPn, with the response scaling as $\epsilon^{1/n}$ for $\epsilon\ll 1$ \cite{chen2017exceptional,wiersig2016sensors,wiersig2020review,zhang2019quantum,wiersig2014enhancing,hokmabadi2019non,mao2023enhanced,zhong2020hierarchical,mandal2021symmetry,sayyad2022realizing}.
This power-law behavior, increasing dramatically with EPs' order $n$, highlights the system's advantage as sensors based on higher-order EPs (HOEPs) in metrology \cite{hodaei2017enhanced,zeng2019enhanced,wu2021high,xiao2019enhanced}.

Despite the promise of EPs, the presence of intrinsic Langevin noise in non-Hermitian systems poses an obstacle to the sensitivity enhancements \cite{langbein2018no,chen2019sensitivity,wang2020petermann,duggan2022limitations,anderson2023clarification,ding2023fundamental,naikoo2023multiparameter,loughlin2024exceptional}.
To solve this issue, several strategies have been proposed, including the use of post-selection techniques to discard quantum noise \cite{naghiloo2019quantum}, and Hamiltonian dilation by integrating the non-Hermitian system and the reservoir as a large Hermitian system \cite{wu2019observation,Sergeev2023signatureof}. 
Recent demonstrations have shown that EPs can also be constructed within Hermitian systems that include nonlinear coupling \cite{wang2019non,jiang2019anti,luo2022quantum}. 
In such systems, the EPs delineate the dynamical phase transition between parametric oscillation and parametric amplification.
Consider a quadratic bosonic system with Hamiltonian:
\begin{equation}
    \hat{\mathcal{H}} = \sum_{i}^{n}\delta_{i}\hat{a}_{i}^{\dagger}\hat{a}_{i}+\sum_{i,j}^{n}(\kappa_{ij}\hat{a}_{i}^{\dagger}\hat{a}_{j}+g_{ij}\hat{a}_{i}^{\dagger}\hat{a}_{j}^{\dagger}+h.c.). \label{Eq.general_bosonic_Hamiltonian}
\end{equation}
The first two terms in (\ref{Eq.general_bosonic_Hamiltonian}) are particle-conserving, manifesting as periodic oscillations in the system's dynamics, while the third term is particle-nonconserving, leading to exponential growth in the dynamical behavior.
EPs appear at the parametric amplification threshold of the system where the exponential growth terms reach a certain equilibrium with the periodic oscillation terms. 
These EPs emerge solely from the system's non-Hermitian-like behaviors, rather than from losses and gains, and are therefore not subject to Langevin noise limitations. 

In the frame of Hamiltonian (\ref{Eq.general_bosonic_Hamiltonian}), we propose a cavity-mediated atomic sensor that exploits HOEPs with nonlinear responses to external perturbations while immunizing Langevin noise.
The cavity mode interacts with the collective spin-wave excitations (\textit{magnons}) through either SU(1,1)- or SU(2)-type Raman interaction \cite{chen2009enhanced,hammerer2010quantum,chen2010observation,chen2015atom,guo2019high,wen2019non}, and establishes couplings among multiple atomic ensembles to form HOEPs.
The HOEPs are anticipated to enhance the sensing performance of atoms in probing various fields in demand, such as electric fields \cite{khadjavi1968stark}, magnetic fields \cite{budker2007optical}, optical fields \cite{delone1999ac}, and even exotic fields \cite{pustelny2013global}.
The ideology considered to construct HOEPs in the work includes two important aspects:
(i) Although the Hamiltonian with Raman interaction in our system is Hermitian, the system's dynamical matrix exhibits non-Hermitian behaviors satisfying pseudo-Hermiticity \cite{mostafazadeh2010pseudo,mostafazadeh2004physical,mostafazadeh2002pseudo} and particle-hole symmetry \cite{delplace2021symmetry,sayyad2022realizing,okugawa2019topological}. 
The spontaneous breaking of these symmetries induces the emergence of HOEPs through tuning the interaction strengths and detunings.
(ii) 
The essential condition for the system to achieve the highest order EP requires the irreducibility of dynamical matrix characterizing the entire atom-cavity system.
This irreducibility can be determined in terms of the Casimir invariants of the SU(1,1) and SU(2) transformations \cite{yurke19862,ban1993lie}. 
As an illustration, an EP3-based atomic sensor is devised, showing sensitivity orders of magnitude superior to EP2-based one. 
This sensitivity enhancement completely arises from the HOEP-induced signal amplification at the given working points.

\begin{figure}[t]
	\centering
	\includegraphics[scale=0.5]{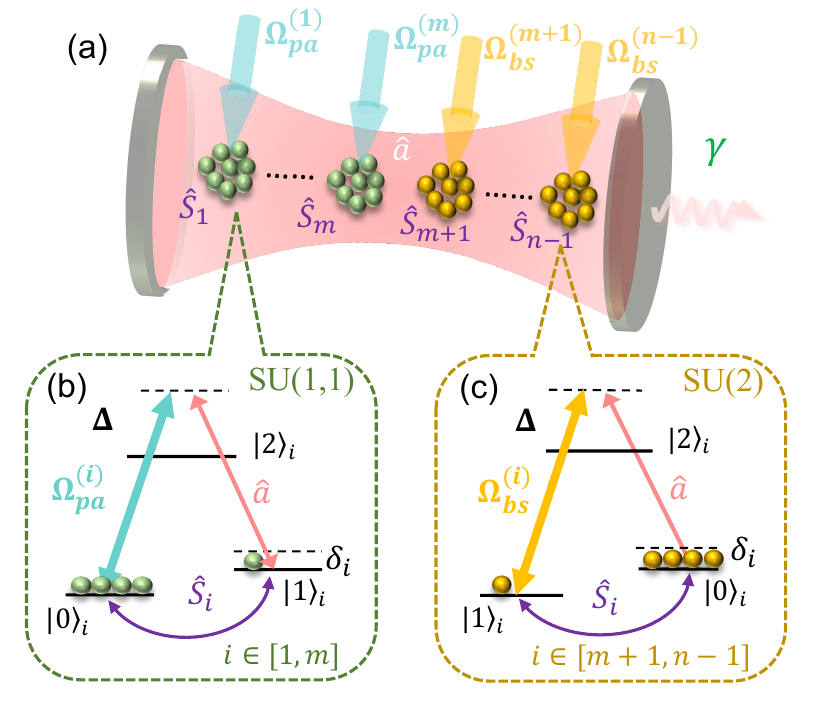}
	\caption{(a) Atom-cavity system to realize EPn. $n-1$ atomic ensembles are placed inside a cavity, where $m$ ($1 \leq m \leq n-1$) ensembles interact with the cavity mode through SU(1,1)-type Raman interactions (b), and the remaining $n-m-1$ ensembles interact with the cavity mode through SU(2)-type Raman interactions (c). (b) and (c) depict the atomic energy levels and mechanisms of the SU(1,1)- and SU(2)-type atom-light interactions, respectively.}
	\label{figure1}
\end{figure}

\textit{The model.}-- The proposed atom-cavity system is illustrated in Fig. \ref{figure1}. 
Each atomic ensemble consists of $N_{a}$ atoms with a pair of lower metastable states $|0\rangle_{i}$ and $|1\rangle_{i}$, as well as an excited state $|2\rangle_{i}$.
Here $i=1,\cdots,n-1$ labels a particular atomic ensemble.  
Coherence between the states $|0\rangle_{i}$ and $|1\rangle_{i}$ can be established and represented by the atomic spin wave$\hat{S}_{i} = \sum_{k}^{N_{a}}|0\rangle_{kk}\langle 1|_{i}$.
Using the Holstein-Primakoff transformation \cite{holstein1940field} and assuming that the mean number of atoms transferred to the state $|1\rangle_{i}$ is much smaller than $N_{a}$, we can view the collective spin-wave excitations in the atomic ensembles as quasiparticles known as \textit{magnons} and express it in terms of bosonic annihilation operator $\hat{b}_{i}$, i.e. $\hat{S}_{i} = (N_{a}-\hat{b}_{i}^{\dagger}\hat{b}_{i})^{1/2} \hat{b}_{i} \approx \sqrt{N_{a}}\hat{b}_{i}$.
As an experimental choice with alkali atoms, e.g., $^{87}\text{Rb}$, the states $|0\rangle$ and $|1\rangle$ correspond to $|5S_{1/2},F=1,m_{F}=\pm 1\rangle$, and $|2\rangle$ corresponds to $|5P_{1/2},F^{'}=1,m_{F^{'}}=0\rangle$.
An external field can be applied to lift the degeneracy of the $m_{F}=\pm 1$ states and enable distinct transition channels between these states. 
The transitions $m_{F}=\mp1$ to $m_{F^{'}}=0$ can be coupled using circularly polarized $\sigma_{+}$ pump and $\sigma_{-}$ cavity fields, respectively.

Two types of Raman interactions, SU(1,1)- and SU(2)-type ones \cite{chen2009enhanced,hammerer2010quantum,chen2010observation,chen2015atom,guo2019high,wen2019non}, are exploited between the cavity field and the magnon fields.
For convenience, we assume that the first $m$ atomic ensembles interact with the cavity mode via SU(1,1)-type Raman interactions.
They are initially prepared through optical pumping in state $|0\rangle_{i}$ with $i\in [1,m]$ [see Fig. \ref{figure1} (b)].
Pump light $\Omega_{pa}^{(i)}$ couples the state $|0\rangle_{i}$ and the excited state $|2\rangle_{i}$ with a large single-photon detuning $\Delta$ to simultaneously generate correlated magnon field $\hat{b}_{i}$ and cavity field $\hat{a}$. 
By adiabatically eliminating the excited state $|2\rangle_{i}$, we have the Hamiltonian $\hat{\mathcal{H}}_{SU(1,1)} \propto \hat{b}_{i}^{\dagger}\hat{a}^{\dagger}+h.c.$ that obeys the Lie algebra of SU(1,1).
The remaining $n-m-1$ atomic ensembles experience the SU(2)-type Raman interactions. 
They are initially prepared through optical pumping in another metastable state, here for convenience, relabelled as $|0\rangle_{i}$ with $i\in[m+1,n-1]$ in our theoretical treatment [see Fig. \ref{figure1} (c)].
The magnon field $\hat{b}_{i}$ interchanges with the cavity field $\hat{a}$ via pump light $\Omega_{bs}^{(i)}$ detuned by $\Delta$. By performing adiabatically elimination of $|2\rangle_{i}$, we end at Hamiltonian $
\hat{\mathcal{H}}_{SU(2)} \propto \hat{b}_{i}^{\dagger}\hat{a}+h.c.$ that obeys the Lie algebra of SU(2).

The Hamiltonian of the whole system is then given by:
\begin{equation}
\hat{\mathcal{H}}  = \hat{\mathcal{H}}_{0}+ \sum_{i=1}^{m} g_{i} (\hat{b}_{i}^{\dagger}\hat{a}^{\dagger}+ \hat{a}\hat{b}_{i}) + \sum_{i=m}^{n-1}\kappa_{i-m} (\hat{b}_{i}^{\dagger}\hat{a}+ \hat{a}^{\dagger}\hat{b}_{i}) , \label{Eq.Hamiltonian_boson}
\end{equation}
where $\hat{\mathcal{H}}_{0} = \sum_{i=1}^{n-1}\hat{\mathcal{H}}_{i}= \sum_{i=1}^{n-1}(\delta_{i}+\epsilon_{i})\hat{b}_{i}^{\dagger}\hat{b}_{i}$ is the free-field Hamiltonian for all magnon modes, $\delta_{i}$ is the two-photon detuning and $\epsilon_{i}$ represents the external perturbation acting on the particular magnon mode $i$, $g_{i} = \sqrt{N_{a}}|\Omega_{pa}^{(i)} g_{pa}|/\Delta$ and $\kappa_{i-m} = \sqrt{N_{a}}|\Omega_{bs}^{(i)} g_{bs}|/\Delta$ (here $\kappa_{0}$ is set to zero) are the effective coupling coefficients of SU(1,1)- and SU(2)-type Raman interaction, respectively. $g_{pa}$ and $g_{bs}$ represent the atomic transition coefficients. 
We utilize such an atom-cavity system as a quantum sensor to detect the external perturbations $\epsilon_{i}$. 
In the following context, we set all atomic ensembles to experience the same perturbation on the detuning $\epsilon_{i}\equiv\epsilon$, The result of $\epsilon_{i}\neq\epsilon_{j}$ is shown in \cite{suppmat}.

\textit{Emergence of higher-order exceptional point.}-- 
Despite the Hermiticity of the system's Hamiltonian, non-conservative SU(1,1) terms $\hat{a}^{\dagger}\hat{b}_{i}^{\dagger}+\hat{b}_{i}\hat{a}$ lead to a non-Hermitian-like dynamics of system.
We can formulate the system's dynamics based on the Heisenberg equation:
\begin{equation}
    i\frac{\partial}{\partial t} \vec{\bm{\Phi}}(t) = H_{D}^{(n,m)} \vec{\bm{\Phi}}(t). \label{Eq.SD}
\end{equation} 
Here $\vec{\bm{\Phi}} = (\hat{b}_{1}, \hat{b}_{1}^{\dagger}, ...,\hat{b}_{n-1},\hat{b}_{n-1}^{\dagger}, \hat{a},\hat{a}^{\dagger})^{T}$ is the field vector and $H_{D}^{(n,m)}$ is the non-Hermitian dynamical matrix of our system.
$H_{D}^{(n,m)}$ inherently follows the particle-hole symmetry, i.e. $\mathcal{C} H_{D}^{(n,m)} \mathcal{C}^{-1}=-H_{D}^{(n,m)}$ with $\mathcal{C} = \bm{I}_{n}\otimes\bm{\sigma_{x}}$ being the conjugate parity operation $\{\hat{c} \leftrightarrow \hat{c}^{\dagger}\}$; as well as the pseudo-Hermiticity, i.e. $\eta H_{D}^{(n,m)} \eta^{-1}=H_{D}^{(n,m)\dagger}$ with $\eta = \bm{I}_{n}\otimes\bm{\sigma_{z}}$ for the parity operation $\{\hat{c}^{\dagger} \leftrightarrow -\hat{c}^{\dagger}\}$. Here, $\hat{c}$ represents the magnon/photon annihilation operator $\{\hat{b}_{i},\hat{a}\}$, $\bm{\sigma_{x,y,z}}$ are the Pauli matrices and $\bm{I}_{n}$ being the $n\times n$ identity matrix. The presence of these symmetries enables the non-Hermitian matrix $H_{D}^{(n,m)}$ to manifest real eigenvalues, which correspond to dynamically stable solutions in the system.
The spontaneous breaking of these symmetries induces a dynamical phase transition, leading to the emergence of EPs of at least order two.

\begin{figure}[t]
	\centering
	\includegraphics[scale=0.42]{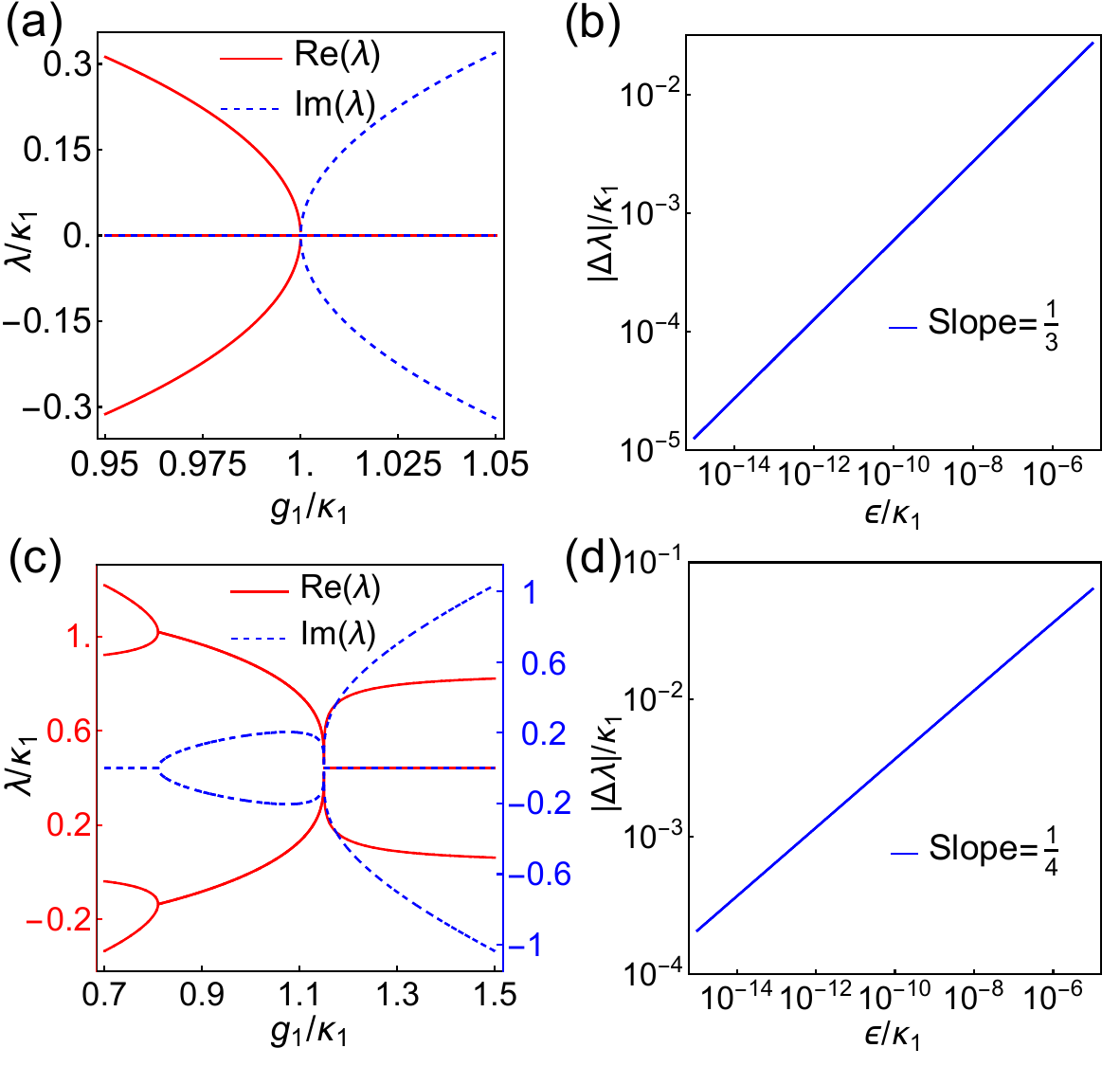}
	\caption{EP3 and EP4 realized in Hermitian atom-cavity system.(a) The eigenvalues of $H_{D}^{(3,1)}$. Red solid (blue dashed) lines represent the real (imaginary) parts. Parameters used are $\delta_{1}=\delta_{2}=0$. (b) Logarithmic scale plot of the absolute eigenvalue splitting at EP3 ($g_{1}=\kappa_{1}$) as a function of perturbation $\epsilon$ with slope=1/3. (c) The eigenvalues of $H_{D}^{(4,2)}$. Parameters used are $\delta_{1}=0.8845\kappa_{1}$, $\delta_{2}=0.0340\kappa_{1}$, $\delta_{3}=-0.8505\kappa_{1}$ and $g_{2}=0.2\kappa_{1}$. (d) Logarithmic scale plot of the absolute eigenvalue splitting at EP4 ($g_{1}=1.1499\kappa_{1}$) as a function of perturbation $\epsilon$ with slope=1/4.}
	\label{Fig.eigenvalue}
\end{figure}

Achieving the highest-order EPn in the atom-cavity system necessitates that $H_{D}^{(n,m)}$ be irreducible, indicating that the system cannot be decomposed into independent subsystems under linear transformations. 
This can be examined in terms of the free magnon Hamiltonians ($\hat{\mathcal{H}}_{i}$ and $\hat{\mathcal{H}}_{j},\, i\neq j$) and the corresponding Casimir invariants \cite{yurke19862,ban1993lie}, i.e., $\hat{b}_{i}^{\dagger}\hat{b}_{i}-\hat{b}_{j}^{\dagger}\hat{b}_{j}$ for SU(1,1)-type interaction and $\hat{b}_{i}^{\dagger}\hat{b}_{i}+\hat{b}_{j}^{\dagger}\hat{b}_{j}$ for SU(2)-type one.
For two magnon modes ($i, j$) engaging in different types of interactions with the cavity mode, i.e., $i \in [1,m]$, $j\in[m+1, n-1]$, $\hat{\mathcal{H}}_{i}+\hat{\mathcal{H}}_{j}$ must not be proportional to the $\hat{b}_{i}^{\dagger}\hat{b}_{i}-\hat{b}_{j}^{\dagger}\hat{b}_{j}$.
For two magnon modes ($i, j$) both engaging in the same type of interactions with the cavity mode, 
i.e., either $i,j \in [1,m]$ or $i,j \in [m+1,n-1]$, 
$\hat{\mathcal{H}}_{i}+\hat{\mathcal{H}}_{j}$ must not be proportional to $\hat{b}_{i}^{\dagger}\hat{b}_{i}+\hat{b}_{j}^{\dagger}\hat{b}_{j}$.
In general, to achieve the highest-order EPn, the two-photon detunings adhere to the condition $\delta_{1}^{'} \neq \cdots \neq \delta_{m}^{'} \neq -\delta_{m+1}^{'}\neq \cdots \neq -\delta_{n-1}^{'}$ where $\delta_{i}^{'}=\delta_{i}+\epsilon_{i}$. 

For illustration, we consider a system with two atomic ensembles. 
The two magnon modes interact with the cavity field through SU(1,1) and SU(2) interactions, respectively.
The dynamical matrix is as follows:
\begin{equation}
    H_{D}^{(3,1)} = \left(\begin{array}{cccc}
		\delta_{1}^{'}\bm{\sigma_{z}} & \, \bm{0} & \, ig_{1}\bm{\sigma_{y}} \\
		\bm{0} & \, \delta_{2}^{'}\bm{\sigma_{z}} & \, \kappa_{1}\bm{\sigma_{z}}  \\
		ig_{1}\bm{\sigma_{y}} & \, \kappa_{1}\bm{\sigma_{z}} & \, \bm{0} 
	\end{array}\right).
 \label{Eq.HD}
\end{equation}
This system can achieve EP3 when the dynamical matrix (\ref{Eq.HD}) is irreducible under the essential condition $\delta_1 = \delta_2 = 0$ and $\epsilon_1 \neq -\epsilon_2$ [see Fig. \ref{Fig.eigenvalue} (a-b)].
At this EP3, the eigenvalues exhibit cubic-root responses to perturbations $\epsilon$ in two-photon detuning.
A reducible case of the matrix (\ref{Eq.HD}), as a \textit{counterexample}, shows that only an EP2 exists (see Supplemental Material \cite{suppmat}).
Further extending this system by adding an atomic ensemble undergoing an SU(1,1)-type interaction with the cavity mode elevates the order of EP to four with the condition $\delta_{1}^{'}\neq\delta_{2}^{'}\neq-\delta_{3}^{'}$ [see Fig. \ref{Fig.eigenvalue} (c-d)].
At this EP4, the eigenvalues exhibit biquadratic-root responses to the perturbations $\epsilon$ in two-photon detuning.

\textit{EP3-based sensor.}-- 
The higher-order response at the HOEP suggests dramatic changes in the system's dynamics, indicating enhanced performance of the HOEP-based sensor.
For demonstration, we develop an EP3-based sensor in the system described by the dynamical matrix (\ref{Eq.HD}).

The system's dynamics on both sides of EP3 exhibit different behaviors: parametric oscillation when $g_{1}<\kappa_{1}$ and parametric amplification when $g_{1}>\kappa_{1}$.
During the parametric-amplification phase, atomic excitation grows exponentially with time, inducing extra higher-order nonlinear effects (such as $\hat{b}_{1}^{\dagger}\hat{b}_{1}\hat{b}_{1}\hat{a}+\hat{b}_{2}^{\dagger}\hat{b}_{2}^{\dagger}\hat{b}_{2}\hat{a}+h.c.$) in atomic ensembles that disrupt the non-Hermitian symmetries.
Therefore, we focus on the EP3-based sensor operating in the parametric-oscillation phase.

For a perfect optical cavity, the time evolution of the magnons can be obtained from Eq. (\ref{Eq.SD}) and Eq. (\ref{Eq.HD}):
\begin{eqnarray}
    \hat{b}_{1}(t) &=& A_{1} \hat{b}_{1}(0) -iB \hat{a}^{\dagger}(0) + C \hat{b}_{2}^{\dagger}(0), \\
    \hat{b}_{2}(t) &=& -C \hat{b}_{1}^{\dagger}(0) -iD \hat{a}(0) + A_{2} \hat{b}_{2}(0). 
\end{eqnarray}
Here, $\{A_{1}, A_{2}, B, C, D\}$ are the propagation coefficients depending on $\{g_{1}, \kappa_{1}, \epsilon, t\}$ which determine the evolution of magnons.
By constructing a biorthogonal complete eigenbasis of the dynamical matrix and applying non-Hermitian perturbation theory \cite{li2023speeding}, we obtain the first-order perturbation approximation for these propagation coefficients that matches well with the numerical results when $\epsilon\ll (\kappa_{1}^{2}-g_{1}^{2})^{3/2}$ \cite{suppmat}.

We can estimate the external perturbations $\epsilon$ by performing homodyne measurements on the magnon modes. This can be implemented by leveraging an SU(2)-type Raman interaction to convert the magnons into photons, which are then detected through standard homodyne measurements \cite{suppmat}.

\begin{figure}[t]
	\centering
	\includegraphics[scale=0.44]{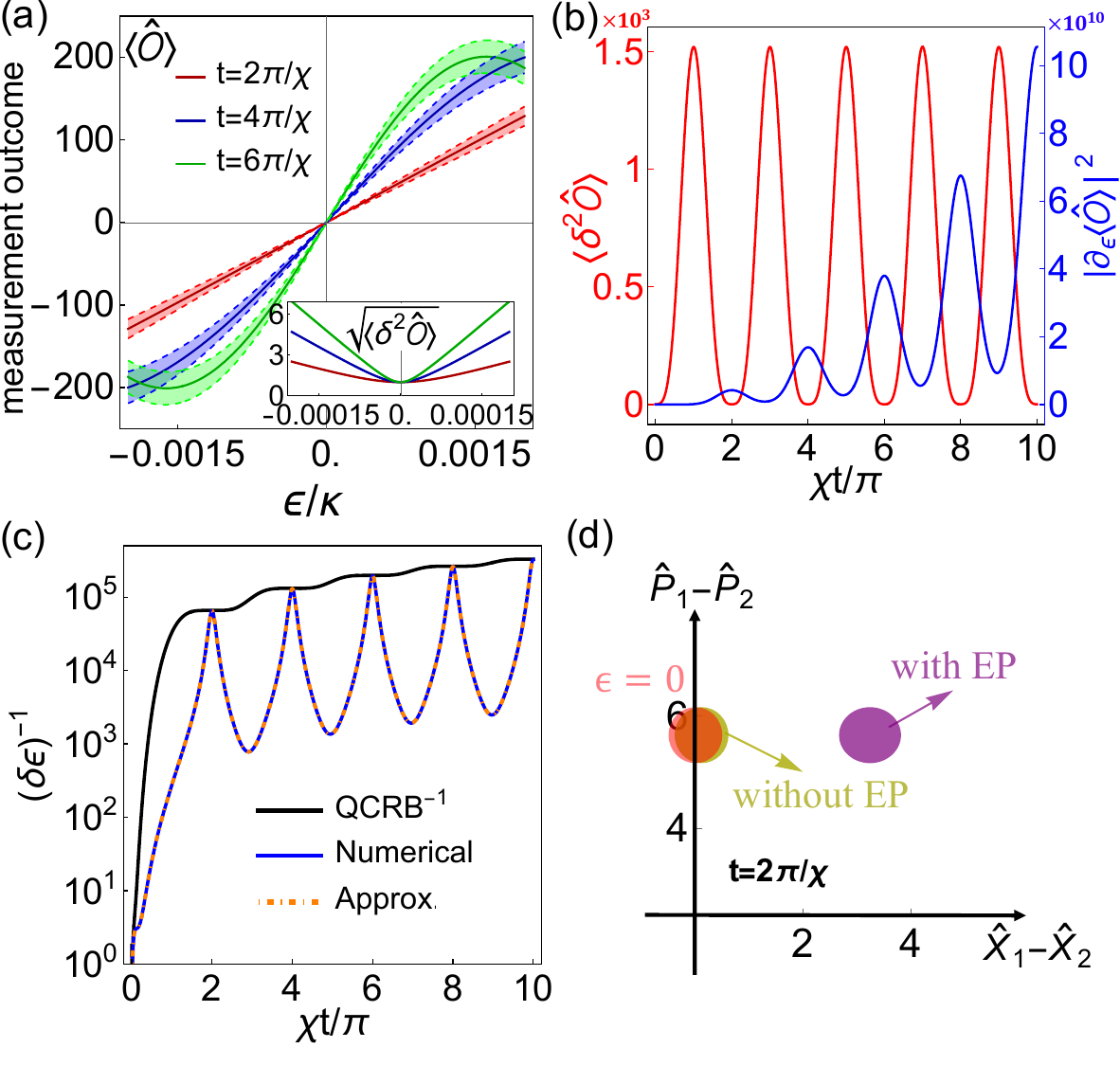}
	\caption{(a) The mean value $\langle\hat{\mathcal{O}}\rangle$ and the standard deviation $\sqrt{\langle\delta^{2}\hat{\mathcal{O}}\rangle}$ (the inset) as a function of $\epsilon$ at different working points. The shaded areas represent the measurement errors. (b) The variance $\langle\delta^{2}\hat{\mathcal{O}}\rangle$ (red) and susceptibility $|\partial_{\epsilon}\langle\hat{\mathcal{O}}\rangle|$ (blue) of observable $\hat{\mathcal{O}}=\hat{X}_{1}-\hat{X}_{2}$ as a function of evolution time when $\epsilon=0$. (c) The reciprocal of sensitivity $(\delta\epsilon)^{-1}$ as a function of evolution time when $\epsilon=0$. The black solid curve represents the QCRB \cite{suppmat}. In (a-c), we set $g/\kappa=0.95$, $\kappa=1$, and $\alpha=2$. (d) The error contours of the magnon state at $\epsilon=0$ (red circle) and at $\epsilon=5\times10^{-5} \kappa$ (purple circle). The result is compared with the case without EP (yellow circle) using the same magnon number $N \sim \chi^{-4}$. In all plots, we set $g/\kappa=0.95$, $\kappa=1$, and $\alpha=2$. }
	\label{figure3}
\end{figure}

We consider a coherent initial state for magnon modes $|\alpha_{1}\rangle_{b_{1}}|\alpha_{2}\rangle_{b_{2}}$.
For convenience, we relabel $g_{1}=g, \kappa_{1}=\kappa$. 
Without loss of generality, we further set $\alpha_{1}=-\alpha_{2}\equiv i\alpha \, (\alpha\in \Bbb{R})$, and then only the quadratures $\hat{X}_{1}$ and $\hat{X}_{2}$ are sensitive to the phase perturbation $\epsilon$. Here, $\hat{X}_{i} = (\hat{b}_{i}+\hat{b}_{i}^{\dagger})/\sqrt{2}$ is the quadrature operator of magnon mode $i$.
First, we consider the observable $\hat{\mathcal{O}}=\hat{X}_{1} - \hat{X}_{2}$ to measure the system's change.
The system's change is quantified by the susceptibility $\mathcal{S}=|\partial_{\epsilon}\langle\hat{\mathcal{O}}\rangle|$, which is maximized near $\epsilon = 0$, as shown in Fig. \ref{figure3} (a).
From the first-order perturbation theory, we have the susceptibility \cite{suppmat}:
\begin{equation}
\mathcal{S}|_{\epsilon=0} =
    \frac{\sqrt{2}\alpha\,(\kappa+g)^{2}\xi}{2\chi^{5}}, \label{Eq.signal}
\end{equation}
where $\xi=(\kappa^{2}+g^{2}) \chi t[2 +\text{cos}(\chi t)]+(g^{2}-8g\kappa+\kappa^{2})\text{sin}(\chi t)$ and $\chi =\sqrt{\kappa^{2}-g^{2}}$. 
The local maxima $\mathcal{S}_{\text{max}}=3\sqrt{2}\alpha(\kappa^{2}+g^{2})(\kappa+g)^{2}q\pi/\chi^{5}\sim \chi^{-5}$ can be reached at $t=2q\pi/\chi$ with $q=1,2,\cdots$ [see blue curve in Fig. \ref{figure3} (b)].
Near EP3 ($\chi\rightarrow 0$), the susceptibility exhibits a divergent feature, induced by EP and representing EP-enhanced response. 
The noise of this sensor is the variance of the observable $\langle\delta^{2}\hat{\mathcal{O}}\rangle=[\kappa-g\text{cos}(\chi t)]^{2}/(\kappa-g)^{2}$. 
As shown by the red curve in Fig. \ref{figure3} (b), the local minima of noise $\langle\delta^{2}\hat{\mathcal{O}}\rangle_{\text{min}}=1$ is reached at $t=2q\pi/\chi$. In such a case, the system evolves back to its initial coherent state.
Thus at the working point $\chi t=2q\pi$, the optimal sensitivity of the system is $\delta\epsilon_{\text{opt}} = \sqrt{\langle\delta^{2}\hat{\mathcal{O}}\rangle_{\text{min}}}/\mathcal{S}_{\text{max}}\sim \chi^{5}$ which saturates the quantum Cramér-Rao bound (QCRB), as shown in Fig. \ref{figure3} (c).
With time evolution, the magnon number oscillates and is bounded by $N\sim\chi^{-4}$ \cite{suppmat}. The time for reaching the working point is $t\sim \chi^{-1}$.
Therefore, the optimal sensitivity $\delta\epsilon_{\text{opt}}\sim\chi^{5}$ scales at the Heisenberg limit $N^{-1}t^{-1}$.

As shown in Fig. \ref{figure3} (a) and (d), at the working point $\chi t = 2q\pi$, 
a small $\epsilon$ causes an amplified displacement of the magnon state in phase space, leading to a dramatic change in the measurement outcome $\langle\hat{\mathcal{O}}\rangle$, while the change of the standard deviation $\sqrt{\langle\delta^{2}\hat{\mathcal{O}}\rangle}$ via $\epsilon$ is negligible.
This indicates that the sensitivity enhancement originates from the 
signal amplification through HOEP, rather than from the reduction of noise through quantum squeezing/entanglement.
This conclusion is further confirmed by the Gaussian decomposition \cite{weedbrook2012gaussian,suppmat}.

It is deserved to emphasize that entanglement is generated between two atomic ensembles \cite{parkins2006unconditional,dalla2013dissipative} which exhibits the anti-squeezing feature of the observable $\hat{\mathcal{O}}=\hat{X}_{1}-\hat{X}_{2}$ or $\hat{P}_{1}+\hat{P}_{2}$and the squeezing feature of the alternative observable $\hat{X}_{1}+\hat{X}_{2}$ or $\hat{P}_{1}-\hat{P}_{2}$ off the working point $\chi t \neq 2q\pi$. 
During the first half of the oscillation period ($\chi t \in [0, \pi]$), the magnon state is squeezed due to the parametric interaction, while the detuning $\epsilon$ induces a displacement in the $X$-direction. In the second half ($\chi t \in [\pi, 2\pi]$), the accumulation of detuning leads to a phase reversal of the parametric interaction. As a result, the displacement is amplified, while the noise evolves back to the initial coherent level \cite{kong2013cancellation,hudelist2014quantum,burd2019quantum,agarwal2022quantifying,colombo2022time}. Such noiseless amplification intensifies as the system approaches the HOEP.
One can also utilize quantum squeezing to enhance sensitivity by measuring $\hat{X}_{1}+\hat{X}_{2}$, which saturates the QCRB at $\chi t = (2q+1)\pi$ \cite{suppmat}.
However, the squeezing-based strategy has an intrinsic disadvantage over the EP-based one when there exist losses in the system and detection \cite{alushi2024optimality}.
In general, the losses of the system and detection themselves and their introduced extra noises degrade the sensitivity.
The squeezing-based strategy relies on quantum squeezing which is fragile due to the loss-induced decoherence. 
In contrast, the EP-based one employs noiseless signal amplification which inherently holds loss tolerance and noise resistance
\cite{hudelist2014quantum,hosten2016quantum,manceau2017detection,du2020quantum,du20222}.
A detailed analysis demonstrates the loss dependence of both strategies in \cite{suppmat}.
The results indicate that the EP-based strategy is evidently a preferred candidate for sensitivity enhancement in the atom-cavity system.

\textit{Higher-order EP enhancement.}-- 
For comparison, we consider a quantum sensor based on Langevin noise-free EP2 \cite{luo2022quantum}, with its optimal sensitivity scales at $\chi^{3}$. This is evidently lower than that of our EP3-based sensor (scales at $\chi^{5}$), though both of them scale at the Heisenberg limit.
The advantage of the EP3 arises from the additional SU(2)-type interaction, which significantly increases the sensing resources and enhances the sensitivity \cite{du2020quantum,du20222}.

We further analyze a general EPn-based sensor with initial coherent states and homodyne detection.
For the parametric-oscillation phase with oscillation period $2\pi/\chi \sim \epsilon^{-1/n}$, through evaluating the derivative of propagation coefficients $\delta\epsilon_{\text{opt}}\sim (\partial_{\epsilon}\mathbf{K}_{jk})^{-1} \sim \epsilon^{2-1/n}$, we can deduce a general scaling law for its optimal sensitivity $\delta\epsilon_{\text{opt}}\sim \chi^{2n-1}$ \cite{suppmat}.
This is attributed to the higher amplification rate of atomic (bosonic) excitations near HOEP, enhancing the sensitivity.

\textit{Experimental feasibility.}--
Here, we provide a possible experimental condition.
We assume $N_{a}\sim 10^{6}$ $^{87}\text{Rb}$ atoms, they are initially prepared in spin coherent states with $\alpha \sim 10^{2}$.
We select atomic transition coefficients of $g_{pa,bs}/2\pi \sim 50 \text{ kHz}$ and detuning of $\Delta_{pa,bs}/2\pi \sim 250 \text{ MHz}$ \cite{kruse2003cold,nagorny2003optical,nagorny2003collective}.
With pump beams of approximately $100\,\mu\text{W}$ and waists around $600 \,\mu\text{m}$, the resulting Rabi frequencies are $\Omega_{pa,bs}/2\pi \sim 2.6 \text{ MHz}$.
This setup can yield magnon-photon interaction strengths of $g/2\pi, \kappa/2\pi \sim 500 \text{ kHz}$. 
Focusing on the EP3-based sensor, we can fine-tune $g$ and $\kappa$ by adjusting the intensity of the pump light. Within the allowed precision of adjustment, $g/\kappa =0.995$ is achievable, corresponding to $\chi \sim 50 \text{ kHz}$.
Given these parameters, this $^{87}\text{Rb}$ atom-based sensor is capable of detecting perturbations \(\epsilon\), with sensitivity $\delta\epsilon \sim 5.27 \times 10^{-6} \text{ Hz}\,\text{Hz}^{-1/2}$, corresponding to, e.g., a sensitivity $\sim0.75 \text{ fT}\,\text{Hz}^{-1/2}$ for magnetic field measurement at the working point $t=2\pi/\chi$.
With the experimentally achievable cavity decay rate $\gamma=0.1\kappa\sim 2\pi\times50\text{kHz}$ and magnon decay rate $\Gamma=0.01\kappa\sim2\pi\times5\text{kHz}$ \cite{kruse2003cold,nagorny2003optical,nagorny2003collective}, this sensor can still surpass standard quantum limit by over $10 \text{ dB}$.

\textit{Conclusion.}--
In summary, we develop a magnon-cavity sensor that allows Langevin noise-free HOEPs to achieve high sensitivity. 
Notably, this sensor demonstrates a sensitivity far surpassing that of EP2-based sensors. 
Due to the versatility of atoms, the external perturbation $\epsilon$ can be induced by different interactions of atoms with various measured fields, including the Stark effect from electric fields \cite{khadjavi1968stark}, the Zeeman effect from magnetic fields \cite{budker2007optical}, the AC Stark effect from optical fields \cite{delone1999ac}, and even exotic interactions from axion-like particles \cite{pustelny2013global}.
Furthermore, compared with other platforms such as superconducting circuits, atom-based ones offer advantages in experimental simplicity without cryogenic systems \cite{degen2017quantum} and the capability to operate in both optical and microwave frequencies.
The crucial contribution of Langevin noise-free HOEPs in enhancing the atom-cavity system's response to various measured fields represents a significant advance in precision measurement, indicating a new direction in sensing with quantum technology.

We thank Prof. Xin-You Lü from Huazhong University of Science and Technology and Prof. Keye Zhang from East China Normal University for useful discussion.
This work is supported by the Innovation Program for Quantum Science and Technology (2021ZD03032001); the National Natural Science Foundation of China (12234014, 11654005, 12204303, 11904227, 12374331); the Shanghai Municipal Science and Technology Major Project (2019SHZDZX01); the National Key Research and Development Program of China (Grant No. 2016YFA0302001); W. Z. also acknowledges additional support from the Shanghai talent program.

\nocite{*}

\bibliography{ref_v3}

\appendix

\newpage

\newpage
\setcounter{equation}{0}
\setcounter{figure}{0}
\renewcommand{\thefigure}{S\arabic{figure}} 
\renewcommand{\theequation}{S\arabic{equation}}

\begin{widetext}
\renewcommand{\thefigure}{S\arabic{figure}} 
\renewcommand{\theequation}{S\arabic{equation}}

In this supplemental material, we provide further information and derivation of the system's eigenvalues, evolution, first-order perturbation approximation, the magnon modes read-out process, quantum fisher information, Gaussian decomposition, the impacts of losses, and the sensitivity scaling of EPn-based sensor, to support the results presented in the main text. 



\section{S1. Third and fourth-order exceptional point}
In order to solve the eigenvalues of the system's dynamical matrix and systemic dynamics, we only need to consider half of the field operators in the field operator vector $\Vec{\Phi}(t)$ in the main text, as the other half are the Hermitian conjugates of this half.
We choose the field operator $\hat{b}_{i}$ ($i\in[1,m]$) for the atomic ensembles undergoing SU(1,1)-type interaction, $\hat{b}_{i}^{\dagger}$ ($i\in[m+1,n-1]$) for the atomic ensembles undergoing SU(2)-type interaction, and the cavity field operator $\hat{a}^{\dagger}$ to form the shortened $n$-dimensional field vector $\Vec{\phi} = (\hat{b}_{1},\cdots,\hat{b}_{m},\hat{b}_{m+1}^{\dagger},\cdots,\hat{b}_{n-1}^{\dagger},\hat{a}^{\dagger})$.
The Heisenberg equation of motion derived from this shortened field vector $\Vec{\phi}$ can fully describe the dynamical behavior of the system.
The corresponding simplified $n\times n$ dynamical matrix is then given by
\begin{equation}
    h_{D}^{(n,m)} = \begin{pmatrix}
        \delta_{1}^{'} &\cdots & \bm{0} & \bm{0} & \cdots & \bm{0} & g_{1} \\
        \vdots & \ddots & \vdots &\vdots  &\ddots &\vdots &\vdots \\
        \bm{0} & \cdots& \delta_{m}^{'} & \bm{0} & \cdots & \bm{0} & g_{m} \\
        \bm{0} & \cdots& \bm{0} &  -\delta_{m+1}^{'} & \cdots & \bm{0} & -\kappa_{1} \\
        \vdots & \ddots & \vdots &\vdots  &\ddots &\vdots &\vdots \\
        \bm{0} & \cdots & \bm{0} & \bm{0} & \cdots & -\delta_{n-1}^{'} & -\kappa_{n-m-1} \\
       -g_{1} & \cdots & -g_{m} & -\kappa_{1} & \cdots & -\kappa_{n-m-1} & 0 
    \end{pmatrix},
    \label{Seq.hdnm}
\end{equation}
with $\delta_{i}^{'}=\delta_{i}+\epsilon_{i}$.
In the following context, we use this shortened field vector $\Vec{\phi}$ and simplified dynamical matrix $h_{D}^{(n,m)}$ to calculate the eigenvalues and the system's dynamics for convenience.

\subsection{A. EP3 in 3-mode system for one SU(1,1) and one SU(2) case}
In this subsection, we calculate the solution of eigenvalues of a 3-mode system that consists of two atomic ensembles and one cavity mode. The cavity mode interacts with one of these two atomic ensembles through SU(1,1)-type Raman interaction, while interacting with the other one through SU(2)-type Raman interaction. 
The corresponding simplified $3\times 3$ dynamical matrix for this system is:
\begin{equation}
    h_{D}^{(3,1)} = \left(\begin{array}{cccc}
		\delta_{1}^{'} & \, 0 & \, g \\
		0 & \, -\delta_{2}^{'} & \, -1  \\
		-g & \, -1 & \, 0  
	\end{array}\right). \label{Seq.hD31}
\end{equation}
Here, we set $g_{1}=g,$ and $\kappa_{1}=1$ for convenience.
The diagonalization of $h_{D}^{(3,1)}$ can be done by the Cardano's method \cite{kurosh1972higher}. Its characteristic equation is given by:
\begin{equation}
    \lambda^{3}-(\delta_{1}^{'}-\delta_{2}^{'})\lambda^{2}+(g^{2}-\delta_{1}^{'}\delta_{2}^{'}-1)\lambda+g^{2}\delta_{2}^{'}+\delta_{1}^{'}=0. \label{Seq.lambda_eq}
\end{equation}
The solution is:
\begin{eqnarray}
    \lambda_{1}&=&\frac{\delta_{1}^{'}-\delta_{2}^{'}}{3}-e^{i\pi/3}Z_{+}-e^{-i\pi/3}Z_{-},\label{Seq.eigenvalues1} \\
    \lambda_{2}&=&\frac{\delta_{1}^{'}-\delta_{2}^{'}}{3}+Z_{+}+Z_{-}, \\
    \lambda_{3}&=&\frac{\delta_{1}^{'}-\delta_{2}^{'}}{3}-e^{-i\pi/3}Z_{+}-e^{i\pi/3}Z_{-},   \label{Seq.eigenvalues3}
\end{eqnarray}
where $Z_{\pm}=(y\pm \sqrt{x^{3}+y^{2}})^{1/3}$ with
\begin{eqnarray}
    x&=&\frac{3g^{2}-3-\delta_{1}^{'2}-\delta_{2}^{'2}-\delta_{1}^{'}\delta_{2}^{'}}{9}, \\
    y&=&\frac{-3\delta_{1}^{'}(3g^{2}+\delta_{2}^{'2}+6)-\delta_{2}^{'}(18g^{2}+2\delta_{2}^{'2}+9)+3\delta_{1}^{'2}\delta_{2}^{'}+2\delta_{1}^{'3}}{54}.  \label{Seq.xy}
\end{eqnarray}

The expression $D=x^{3}+y^{2}$ is called the discriminant of the eigenvalue equation. If $D>0$, then one of the eigenvalues $\lambda_{i}$ is real and other two are complex conjugates $\lambda_{j}=\lambda_{k}^{*}\  (i\neq j\neq k)$. In this case, the system is unstable and diverges in long time limit. If $D<0$, then all eigenvalues are real and unequal, and the system is dynamically stable. The exceptional points (EPs) appear when $D=0$, in which case, at least two eigenvalues are equal. In particular, when $x=y=0$, all three eigenvalues coalesce to form a third-order EP (EP3).


We focus on finding EP3 in the absence of external perturbation ($\epsilon_{1}=\epsilon_{2}=0$). 
When $\delta_{1}=\delta_{2}=0$ and $g=1$, EP3 is achieved with degenerate eigenvalues $\lambda_{1}=\lambda_{2}=\lambda_{3}=0$ and coalesced eigenvectors $(-1,0,1)^{T}$.

Next, we examine the splitting of eigenvalues at this EP3 under the external perturbations. We consider two cases: the same perturbation for both ensembles ($\epsilon_{1}=\epsilon_{2}$) and different perturbations for each ensemble ($\epsilon_{1}\neq\epsilon_{2}$).

\textbf{The same perturbations for both atomic ensembles $\epsilon_{1}=\epsilon_{2}$:}
In this case, we set $\epsilon_{1}=\epsilon_{2} \equiv -\epsilon$. The Eq. \ref{Seq.lambda_eq} then reduces to
\begin{equation}
    \lambda^{3}-\epsilon^2\lambda-2\epsilon=0, \label{Seq.distributed_lambda_eq}
\end{equation}
which can be perturbatively expanded using Newton–Puiseux series \cite{hodaei2017enhanced}. For $\epsilon\ll 1$, we consider the first two terms only: $\lambda\approx c_{1}\epsilon^{1/3}+c_{2}\epsilon^{2/3}$, with the coefficients $c_{1}$ and $c_{2}$ being complex constants. Then Eq. \ref{Seq.distributed_lambda_eq} becomes 
\begin{equation}            (c_{1}^{3}-2)\epsilon+3c_{1}^{2}c_{2}\epsilon^{4/3}+3c_{1}c_{2}^{2}\epsilon^{5/3}+c_{2}\epsilon^{2}-c_{1}\epsilon^{7/3}-c_{2}\epsilon^{8/3}=0. \label{Seq.distributed_NPseries}
\end{equation}
Forcing the coefficients of first two terms to be zero, we obtain: $c_{1}=\{2^{1/3}, e^{-i2\pi/3}2^{1/3}, e^{i2\pi/3}2^{1/3}\}$ and $c_{2}=0$. Therefore we have the perturbed eigenvalues:
\begin{eqnarray}
     \lambda_{1,\epsilon_{1}=\epsilon_{2}}^{EP}(\epsilon)&=&e^{-i\frac{2\pi}{3}}(2\epsilon)^{\frac{1}{3}},\\
    \lambda_{2,\epsilon_{1}=\epsilon_{2}}^{EP}(\epsilon)&=&(2\epsilon)^{\frac{1}{3}},\\
    \lambda_{3,\epsilon_{1}=\epsilon_{2}}^{EP}(\epsilon)&=&e^{i\frac{2\pi}{3}}(2\epsilon)^{\frac{1}{3}}. \label{Seq.distributed_eigenvalues}
\end{eqnarray}

\textbf{Different perturbations for each atomic ensemble $\epsilon_{1}\neq\epsilon_{2}$:} 
In this case, we set $\epsilon_{2}=0, \epsilon_{1}\equiv-\epsilon$ for convenience. Again using Newton–Puiseux series, we have:
\begin{equation}
    (c_{1}^{3}-1)\epsilon+3c_{1}^{2}c_{2}\epsilon^{4/3}+(3c_{1}c_{2}^{2}+c_{2}^{2})\epsilon^{5/3}+(c_{2}^{3}+2c_{1}c_{2})\epsilon^{2}+c_{2}^{2}\epsilon^{7/3}=0. \label{Seq.gradient_NPseries}
\end{equation}
Forcing the coefficients of first two terms to be zero, we obtain the solution: $c_{1}=\{1, e^{-i2\pi/3}, e^{i2\pi/3}\}$ and $c_{2}=0$. The corresponding eigenvalues are:
\begin{eqnarray}
    \lambda_{1,\epsilon_{1}\neq\epsilon_{2}}^{EP}(\epsilon)&=&e^{-i\frac{2\pi}{3}}\epsilon^{\frac{1}{3}},\\
    \lambda_{2,\epsilon_{1}\neq\epsilon_{2}}^{EP}(\epsilon)&=&\epsilon^{\frac{1}{3}},\\
    \lambda_{3,\epsilon_{1}\neq\epsilon_{2}}^{EP}(\epsilon)&=&e^{i\frac{2\pi}{3}}\epsilon^{\frac{1}{3}}. \label{Seq.gradient_eigenvalues}
\end{eqnarray}

It is obvious that in both situations, the eigenvalues respond nonlinearly and scale at $\epsilon^{1/3}$. When $\epsilon_{1}=\epsilon_{2}$, there is a higher response with a factor of $2^{1/3}$. This is attributed to the double signal acquisition in this case, while only one atomic ensemble senses the signal from external perturbation.

\textbf{Counterexample with reducibility:} 
Notice that, if \(\epsilon_{1} = -\epsilon_{2} \equiv \epsilon\) (which violates the essential condition in the main text), we can perform an SU(1,1) transformation to introduce two new effective magnon modes: 
\begin{equation}
    \hat{c}_{1} = \frac{g\hat{b}_{1}+\hat{b}_{2}^{\dagger}}{\sqrt{g^{2}-1}} \quad \text{and} \quad \hat{c}_{2} = \frac{g\hat{b}_{2}+\hat{b}_{1}^{\dagger}}{\sqrt{g^{2}-1}},
\end{equation}
it guarantees the particle number difference $\hat{b}_{1}^{\dagger}\hat{b}_{1}-\hat{b}_{2}^{\dagger}\hat{b}_{2}=\hat{c}_{1}^{\dagger}\hat{c}_{1}-\hat{c}_{2}^{\dagger}\hat{c}_{2}$ being invariant. 
With these new modes, the Hamiltonian of the interaction reduces to $\hat{\mathcal{H}} = \epsilon(\hat{c}_{1}^{\dagger}\hat{c}_{1}-\hat{c}_{2}^{\dagger}\hat{c}_{2}) + \sqrt{g^{2}-1}(\hat{a}^{\dagger}\hat{c}_{1}^{\dagger}+\hat{c}_{1}\hat{a})$ with only one effective magnon mode interacting with the cavity field.
The eigenvalues of the system then become $\lambda_{0}=\epsilon$, $\lambda_{\pm}=(\epsilon\pm\sqrt{4+\epsilon^{2}-4g^{2}})/2$ which only exhibits EP2.


\subsection{B. EP4 in 4-mode system for two SU(1,1) and one SU(2) case}
In this subsection, we derive the eigenvalues of a 4-mode system that is composed of three atomic ensembles and one cavity mode. The cavity mode interacts with the first two atomic ensembles through SU(1,1)-type Raman interaction, while interacting with the other one through SU(2)-type Raman interaction. The corresponding simplified $4\times 4$ dynamical matrix for this system is expressed by:
\begin{equation}
    h_{D}^{(4,2)} = \left(\begin{array}{cccc}
		\delta_{1}^{'} & \, 0 & \, 0 &\, g \\
		0 & \, \delta_{2}^{'} & \, 0 & \, f\\
        0 & \, 0 & \, -\delta_{3}^{'} & \, -1\\
		-g & \, -f & \, -1  & \, 0
	\end{array}\right). \label{Seq.hD42}
\end{equation}
Here, we set $g_{1}=g, g_{2}=f$, and $\kappa_{1}=1$ for convenience.
We focus on finding EP4 in this system in the absence of external perturbation ($\epsilon_{1}=\epsilon_{2}=\epsilon_{3}=0$).
Then, the corresponding characteristic equation is:
\begin{equation}
\begin{aligned}
    \lambda^{4}-&(\delta_{1}+\delta_{2}-\delta_{3})\lambda^{3}+(g^{2}+f^{2}-1+\delta_{1}\delta_{2}-\delta_{1}\delta_{3}-\delta_{2}\delta_{3})\lambda^{2} \\
    &+[f^{2}(\delta_{3}-\delta_{1})+g^{2}(\delta_{3}-\delta_{2})+\delta_{1}+\delta_{2}+\delta_{1}\delta_{2}\delta_{3}]\lambda-g^{2}\delta_{2}\delta_{3}-f^{2}\delta_{1}\delta_{3}-\delta_{1}\delta_{2}=0.
\end{aligned} \label{Seq.lambda_hep4}
\end{equation}
By converting the Eq. \ref{Seq.lambda_hep4} into biquadratic form, we obtain the conditions for the system to have an EP4:
\begin{eqnarray}
    \delta_{1} &=& \frac{4f(1+f^{2})}{(1-f^{2})^{3/2}},\\
    \delta_{2} &=& \frac{4f^{3}}{(1-f^{2})^{3/2}},\\
    \delta_{3} &=& -\frac{4f}{(1-f^{2})^{3/2}},
\end{eqnarray}
with $0<f<1$. 
The eigenvalues are then expressed by:
\begin{equation}
\begin{aligned}
    &\lambda_{1,2,3,4} = \frac{2f(1+f^{2})}{(1-f^{2})^{3/2}}\\
    \pm&\frac{\sqrt{(1+f^{2})^{4}-g^{2}(1-f^{2})^{3}\pm\sqrt{[(1+f^{2})^{4}-g^{2}(1-f^{2})^{3}][(f^{4}-6f^{2}+1)^{2}-g^{2}(1-f^{2})^{3}]}}}{\sqrt{2}(1-f^{2})^{3/2}}.
\end{aligned}
\end{equation}
It is readily to find that EP4 is achieved when $g=(1+f^{2})^{2}/(1-f^{2})^{3/2}$. At this EP4, $\lambda_{1}=\lambda_{2}=\lambda_{3}=\lambda_{4}=2f(1+f^{2})/(1-f^{2})^{3/2}$ and the system has only one coalesced eigenvector $\{-(1+f)^{2}, f\sqrt{1-f^{2}}, \sqrt{1-f^{2}}, 2f\}^{T}$.
Similarly, the eigenvalues' biquadratic-root response at EP4 can be examined by the Newton-Puiseux series.
This will not be reiterated here.

\section{S2. System evolution}
In this section, we provide the solution of the system's dynamics in terms of the evolution matrix $\mathbf{K}$.

\subsection{A. General solution of $n$-mode system}
By diagonalizing $h_{D}^{(n,m)}$ (Eq. \ref{Seq.hdnm}), we obtain a series of eigenvectors $\{\Vec{v}_{i}\}$ ($i=1,\cdots,n$) that are expressed by their corresponding eigenvalues $\{\lambda_{i}\}$:
\begin{equation}
\begin{aligned}
    \Vec{v}_{i} =( -\lambda_{i}&+\sum_{p=2}^{m}\frac{g_{p}^{2}}{\delta_{p}^{'}-\lambda_{i}}+\sum_{p=m+1}^{n-1}\frac{\kappa_{p}^{2}}{\delta_{p}^{'}+\lambda_{i}},\quad \frac{g_{1}g_{2}}{\delta_{2}^{'}-\lambda_{i}},\quad\cdots\quad,\\
    &\quad\frac{g_{1}g_{m}}{\delta_{m}^{'}-\lambda_{i}}, \quad-\frac{g_{1}\kappa_{1}}{\delta_{m+1}^{'}+\lambda_{i}},\quad\cdots\quad,\quad-\frac{g_{1}\kappa_{n-m-1}}{\delta_{n-1}^{'}+\lambda_{i}},\quad g_{1} )^{T},
\end{aligned}
\end{equation}

With these unnormalized eigenvectors $\{\Vec{v}_{i}\}$, we can derive the evolution matrix $\mathbf{K}=\mathbf{\tilde{U}} \text{diag}(e^{-i\lambda_{1}t}, \\ \cdots, e^{-i\lambda_{n}t}) \mathbf{\tilde{U}}^{-1}$, where $\mathbf{\tilde{U}}=(\Vec{v}_{1}, \cdots, \Vec{v}_{n})$ is the unnormalized eigenvector matrix.
The matrix elements of $\mathbf{K}$ represent the propagation coefficients of the system:
\begin{eqnarray}
    \mathbf{K}_{jj} &=& \sum_{p=1}^{n} e^{-i\lambda_{p}t} \frac{F_{n-1}^{(jj)}(\lambda_{p})}{\prod_{q\neq p}^{n}(\lambda_{p}-\lambda_{q})}, \label{Seq.Kelements_jj}\\
    |\mathbf{K}_{jn}| &=& |\mathbf{K}_{nj}|= |\sum_{p=1}^{n} e^{-i\lambda_{p}t} \frac{F_{n-2}^{(jn)}(\lambda_{p})}{\prod_{q\neq p}^{n}(\lambda_{p}-\lambda_{q})}|, \quad\quad  (j\neq n) \label{Seq.Kelements_jn}\\
    \mathbf{K}_{jk} &=& \sum_{p=1}^{n} e^{-i\lambda_{p}t} \frac{F_{n-3}^{(jk)}(\lambda_{p})}{\prod_{q\neq p}^{n}(\lambda_{p}-\lambda_{q})}, \quad\quad\quad\quad\quad\quad (j,k\neq n)
     \label{Seq.Kelements_jk}
\end{eqnarray}
where $F_{n}^{(jk)}(\lambda) = a_{n}^{(jk)}\lambda^{n} + a_{n-1}^{(jk)}\lambda^{n-1} + \cdots + a_{0}^{(jk)}$ is the $n$-th order polynomial for $\lambda$ with $a_{0}^{(jk)}\neq 0$.

\subsection{B. Evolution matrix for 3-mode system described by $h_{D}^{(3,1)}$}
In this subsection, we derive the evolution matrix $\mathbf{K}$ for a 3-mode system that is described by $h_{D}^{(3,1)}$ (Eq. \ref{Seq.hD31}).
In the main text, we approach the EP3 by adjusting $g$, and observing the system's response to external perturbations $\epsilon_{i}$. 
Therefore, we set $\delta_{1}=\delta_{2}=0$ and formulate the evolution matrix $\mathbf{K}$ as a function of $\epsilon_{1},\epsilon_{2}$ and $g$.

With the diagonalization of $h_{D}^{(3,1)}$, we obtain the evolution matrix $\mathbf{K}$ of the equation through its eigenvalues and eigenvectors, which is given by:
\begin{equation}
    \left(\begin{array}{c}
                    \hat{b}_{1}(t) \\
                     \hat{b}_{2}^{\dagger}(t) \\
                     \hat{a}^{\dagger}(t)
                    \end{array} \right) = \mathbf{K} \left(\begin{array}{c}
                    \hat{b}_{1}(0) \\
                     \hat{b}_{2}^{\dagger}(0) \\
                     \hat{a}^{\dagger}(0)
                    \end{array} \right) 
        = \mathbf{\tilde{U}} \left(\begin{array}{cccc}
		e^{-i\lambda_{1}t} & \, 0 & \, 0 \\
		0 & \, e^{-i\lambda_{2}t} & \, 0  \\
		0 & \, 0 & \, e^{-i\lambda_{3}t}  
	\end{array}\right) \mathbf{\tilde{U}}^{-1}\left(\begin{array}{c}
                    \hat{b}_{1}(0) \\
                     \hat{b}_{2}^{\dagger}(0) \\
                     \hat{a}^{\dagger}(0)
                    \end{array} \right).  \label{Seq.evolution}
\end{equation}
Here, $\mathbf{\tilde{U}}$ is the unnormalized eigenvector matrix:
\begin{equation}
        \mathbf{\tilde{U}} = \left(\begin{array}{cccc}
		-\frac{\lambda_{1}^{2}+\epsilon_{2}\lambda_{1}-1}{\epsilon_{2}+\lambda_{1}} & \, -\frac{\lambda_{1}^{2}+\epsilon_{2}\lambda_{2}-1}{\epsilon_{2}+\lambda_{2}} & \, -\frac{\lambda_{1}^{2}+\epsilon_{2}\lambda_{3}-1}{\epsilon_{2}+\lambda_{3}} \\
		-\frac{g}{\epsilon_{2}+\lambda_{1}} & \, -\frac{g}{\epsilon_{2}+\lambda_{2}} & \, -\frac{g}{\epsilon_{2}+\lambda_{3}}  \\
		g & \, g & \, g  
	\end{array}\right).
 \label{Seq.eigenvector}
\end{equation}
From the structure of equations \ref{Seq.lambda_eq}-\ref{Seq.eigenvalues3}, we derive the following relationships between eigenvalues:
\begin{equation}
\left\{
    \begin{aligned}
        &\sum_{i=1}^{3} \lambda_{i} = \epsilon_{1}-\epsilon_{2},  \\
        &\prod_{i=1}^{3} \lambda_{i} = -\epsilon_{1}-g^{2}\epsilon_{2}, \\
        &\prod_{i=1}^{3} (\epsilon_{1}-\lambda_{i}) = g^{2}(\epsilon_{1}+\epsilon_{2}), \\
    \end{aligned}\right.\quad
    \left\{
    \begin{aligned}
        &\prod_{i=1}^{3} (\epsilon_{2}+\lambda_{i}) = -(\epsilon_{1}+\epsilon_{2}), \\
        &\prod_{i=1}^{3} (\lambda_{i}^{2}+\epsilon_{2}\lambda_{i}-1) = -g^{4}. \\
        &\frac{\lambda_{i}^{2}+\epsilon_{2}\lambda_{i}-1}{g^{2}} = \frac{\epsilon_{2}+\lambda_{i}}{\epsilon_{1}-\lambda_{i}}.
    \end{aligned}\right.
\end{equation}
Using these relations, we solve the evolution matrix:
\begin{equation}
    \mathbf{K} = \left(\begin{array}{cccc}
		A_{1} & \, C & \, -iB \\
		-C & \, A_{2} & \, iD  \\
		iB & \, iD & \, A_{a}
        \end{array}\right),
\end{equation}
with propagation coefficients
\begin{eqnarray}
        A_{1} &=& \Omega\sum_{i=1}^{3} e^{-i\lambda_{i}t}(\lambda_{i}^{2}+\epsilon_{2}\lambda_{i}-1)(\lambda_{i+1}-\lambda_{i-1}), \label{Seq.propargating_A1}\\
        A_{a} &=& \Omega\sum_{i=1}^{3} e^{-i\lambda_{i}t}(\lambda_{i}-\epsilon_{1})(\lambda_{i}+\epsilon_{2})(\lambda_{i+1}-\lambda_{i-1}),\\
        A_{2} &=& \Omega\sum_{i=1}^{3} e^{-i\lambda_{i}t}(\lambda_{i}^{2}-\epsilon_{1}\lambda_{i}+g^{2})(\lambda_{i+1}-\lambda_{i-1}), \\
        B &=& ig\Omega\sum_{i=1}^{3} e^{-i\lambda_{i}t}(\epsilon_{2}+\lambda_{i})(\lambda_{i+1}-\lambda_{i-1}), \\
        C &=& -g\Omega\sum_{i=1}^{3} e^{-i\lambda_{i}t}(\lambda_{i+1}-\lambda_{i-1}), \\
        D &=& -i\Omega\sum_{i=1}^{3} e^{-i\lambda_{i}t}(\epsilon_{1}-\lambda_{i})(\lambda_{i+1}-\lambda_{i-1}).\label{Seq.propargating_D}
\end{eqnarray}
Here, $\Omega^{-1}=(\lambda_{1}-\lambda_{2})(\lambda_{2}-\lambda_{3})(\lambda_{1}-\lambda_{3})$ is the normalization factor and $\lambda_{0}=\lambda_{3}, \lambda_{4}=\lambda_{1}$.

\section{S3. First-order perturbation theory for EP3-based sensor}
Differentiating the propagation coefficients in Eq. \ref{Seq.propargating_A1}-\ref{Seq.propargating_D} directly with respect to $\epsilon_{1}$ or $\epsilon_{2}$ is extremely difficult, preventing us from intuitively understanding the result. Therefore, in this section, we use first-order perturbation theory to obtain an approximate solution for the system's dynamics \cite{li2023speeding}.

In non-Hermitian systems, the orthogonality of eigenvectors is disrupted, especially at the EPs, where two or more eigenvectors coalesce, leading to a reduction in the dimension of the Hilbert space $\mathscr{H}$. To address this issue, it is necessary to establish a biorthogonal eigenbasis on the dual Hilbert space $\mathscr{H}^{*}$. The results help us in performing the first-order perturbation approximation.

To perform first-order perturbation approximation near EP3, we first set $\delta_{1}=\delta_{2}=0$ and then construct the unperturbed biorthogonal eigenbasis as a function of $g_{1}$.
In the following, we focus on the stable side of the system dynamics ($g_{1}<\kappa_{1}$). The reason is presented in the main text.
For convenience, we set $\kappa_{1}=1$ and $ g_{1}=g$.

\subsection{A. Biorthogonal eigenbasis of unperturbed $h_{D}^{(3,1)}$}

In the absence of perturbation ($\epsilon_{1}=\epsilon_{2}=0$) but considering internal loss ($\gamma \neq 0$, $\Gamma \neq 0$), the dynamical matrix reads
\begin{equation}
    h_{D}^{(3,1)} = \left(\begin{array}{cccc}
		-i\Gamma & \, 0 & \, g \\
		0 & \, -i\Gamma & \, -1  \\
		-g & \, -1 & \, -i\gamma  
	\end{array}\right), \label{Seq.HD}
\end{equation}
with eigenvalues 
\begin{equation}
    \lambda_{0}=-i\Gamma, \quad \lambda_{\pm}=-\frac{i\gamma_{+}}{2}\pm\chi, \label{Seq.eigenvalue_unpertubed}
\end{equation}
where $\chi = \sqrt{1-g^{2}-\gamma_{-}^{2}/4}$. Here, $\gamma_{\pm}=\gamma\pm\Gamma$. Their corresponding normalized right eigenvectors (i.e., $h_{D}^{(3,1)}|\tilde{R}_{i}\rangle=\lambda_{i}|\tilde{R}_{i}\rangle$) are
\begin{equation}
    |\tilde{R}_{0}\rangle=\sqrt{\frac{1}{1+g^{2}}}\left(\begin{array}{c}
                    1 \\
                     -g \\
                     0
                    \end{array} \right), \quad |\tilde{R}_{\pm}\rangle=\sqrt{\frac{1}{2}}\left(\begin{array}{c}
                    g \\
                    -1 \\
                    -\frac{i\gamma_{-}}{2}\pm\chi
                    \end{array} \right). \label{Seq.rightvector}
\end{equation}
Note that in the case when $\gamma=\Gamma=0$ and $g=1$, three eigenvalues and eigenvectors coalesces with each other ($\lambda_{0}=\lambda_{+}=\lambda_{-}, |\tilde{R}_{0}\rangle=|\tilde{R}_{+}\rangle=|\tilde{R}_{-}\rangle$) which is the feature of EP3. These eigenvectors are not orthogonal:
\begin{equation}
    \langle \tilde{R}_{0}|\tilde{R}_{\pm}\rangle \neq 0, \quad \langle \tilde{R}_{+}|\tilde{R}_{-}\rangle \neq 0. \label{Seq.rightorthogonal}
\end{equation}
This is originated from the non-Hermiticity of $h_{D}^{(3,1)}$.
We thus extend our eigenbasis to the state in the dual space $\mathscr{H}^{*}$, i.e., $h_{D}^{(3,1)\dagger}|\tilde{L}_{i}\rangle=\lambda_{i}^{*}|\tilde{L}_{i}\rangle$.
The corresponding normalized left eigenvectors are:
\begin{equation}
    |\tilde{L}_{0}\rangle=\sqrt{\frac{1}{1+g^{2}}}\left(\begin{array}{c}
                    1 \\
                     g \\
                     0
                    \end{array} \right), \quad |\tilde{L}_{\pm}\rangle=\sqrt{\frac{1}{2}}\left(\begin{array}{c}
                    -g \\
                    -1 \\
                    \frac{i\gamma_{-}}{2}\pm\chi
                    \end{array} \right). \label{Seq.leftvector}
\end{equation}
Similarly, the left eigenvectors are also not orthogonal:
\begin{equation}
    \langle \tilde{L}_{0}|\tilde{L}_{\pm}\rangle \neq 0, \quad \langle \tilde{L}_{+}|\tilde{L}_{-}\rangle \neq 0. \label{Seq.leftorthogonal}
\end{equation}

Indeed, the right and left eigenvectors are non-orthogonal in their respective Hilbert spaces ($\mathscr{H}$ and $\mathscr{H}^{*}$). They however form a biorthogonal eigenbasis in the Liouville space ($\mathscr{L}=\mathscr{H}\otimes\mathscr{H}^{*}$):
\begin{equation}
    \langle \tilde{L}_{0}|\tilde{R}_{\pm}\rangle = \langle \tilde{L}_{\pm}|\tilde{R}_{0}\rangle=\langle \tilde{L}_{-}|\tilde{R}_{+}\rangle = \langle \tilde{L}_{+}|\tilde{R}_{-}\rangle = 0. \label{Seq.biorthogonal}
\end{equation}
Therefore, we obtain the normalized biorthogonal eigenbasis $|R_{0,\pm}\rangle=|\tilde{R}_{0,\pm}\rangle/\sqrt{\langle\tilde{L}_{0,\pm}|\tilde{R}_{0,\pm}\rangle}$ and $|L_{0,\pm}\rangle=|\tilde{L}_{0,\pm}\rangle/\sqrt{\langle\tilde{L}_{0,\pm}|\tilde{R}_{0,\pm}\rangle}$:
\begin{equation}
    |R_{0}\rangle=\sqrt{\frac{1}{\chi^{2}+\frac{\gamma_{-}^{2}}{4}}}\left(\begin{array}{c}
                    1 \\
                     -g \\
                     0
                    \end{array} \right), \quad |R_{\pm}\rangle=\sqrt{\frac{1}{2\chi(\chi\mp \frac{i\gamma_{-}}{2})}}\left(\begin{array}{c}
                    g \\
                    -1 \\
                    -\frac{i\gamma_{-}}{2}\pm\chi
                    \end{array} \right), \label{Seq.renormalized_right_bieigenbasis}
\end{equation}
and
\begin{equation}
    |L_{0}\rangle=\sqrt{\frac{1}{\chi^{2}+\frac{\gamma_{-}^{2}}{4}}}\left(\begin{array}{c}
                    1 \\
                     g \\
                     0
                    \end{array} \right), \quad |L_{\pm}\rangle=\sqrt{\frac{1}{2\chi(\chi\pm \frac{i\gamma_{-}}{2})}}\left(\begin{array}{c}
                    -g \\
                    -1 \\
                     \frac{i\gamma_{-}}{2}\pm\chi
                    \end{array} \right). \label{Seq.renormalized_left_bieigenbasis}
\end{equation}
This biorthogonal eigenbasis also satisfies the normalization and completeness relations:
\begin{equation}
        \langle L_{0,\pm}|R_{0,\pm}\rangle = 1, \quad
        \sum_{i=0,\pm} |R_{i}\rangle\langle L_{i}| = \mathbf{I}. \label{Seq.nor_complete}
\end{equation}

\subsection{B. First-order perturbation approximation}
In this section, we consider the case of existing a small external perturbation ($\epsilon_{1}, \epsilon_{2} \ll \chi^{3}$), and perform first-order perturbation approximation to solve the perturbed system's dynamics.
Given the perturbation Hamiltonian $\hat{\mathcal{H}_{p}}=\epsilon_{1}\hat{b}_{1}^{\dagger}\hat{b}_{1}+\epsilon_{2}\hat{b}_{2}^{\dagger}\hat{b}_{2}$, we have the perturbation part of the dynamical matrix:
\begin{equation}
    H_{p} = \left(\begin{array}{cccc}
		\epsilon_{1} & \, 0 & \, 0 \\
		0 & \, -\epsilon_{2} & \, 0  \\
		0 & \, 0 & \, 0  
	\end{array}\right). \label{Seq.HD_perturbation}
\end{equation}
The perturbation matrix in the space formed by right- and left-eigenvectors ($|R_{0,\pm}\rangle$ and $\langle L_{0,\pm}|$) is then expressed by:
\begin{equation}
\begin{aligned}
     [H_{p}] =& \left(\begin{array}{cccc}
	\langle L_{0}|H_{p}|R_{0}\rangle & \, \langle L_{0}|H_{p}|R_{-}\rangle & \, \langle L_{0}|H_{p}|R_{+}\rangle \\
	\langle L_{-}|H_{p}|R_{0}\rangle & \, \langle L_{-}|H_{p}|R_{-}\rangle & \, \langle L_{-}|H_{p}|R_{+}\rangle  \\
	\langle L_{+}|H_{p}|R_{0}\rangle & \, \langle L_{+}|H_{p}|R_{-}\rangle & \, \langle L_{+}|H_{p}|R_{+}\rangle  
	\end{array}\right) \\
 = &\left(\begin{array}{cccc}
	\frac{\epsilon_{1}+\epsilon_{2} g^{2}}{1-g^{2}} & \, \frac{(\epsilon_{1}+\epsilon_{2})g}{1-g^{2}}\sqrt{\frac{\chi-i\gamma_{-}/2}{2\chi}} & \, \frac{(\epsilon_{1}+\epsilon_{2})g}{1-g^{2}}\sqrt{\frac{\chi+i\gamma_{-}/2}{2\chi}} \\
	\frac{-(\epsilon_{1}+\epsilon_{2})g}{1-g^{2}}\sqrt{\frac{\chi-i\gamma_{-}/2}{2\chi}} & \, \frac{-(\epsilon_{1} g^{2}+\epsilon_{2})}{2\chi(\chi+i\gamma_{-}/2)} & \, \frac{-(\epsilon_{1} g^{2}+\epsilon_{2})}{2\chi}\sqrt{\frac{1}{1-g^{2}}}  \\
	\frac{-(\epsilon_{1}+\epsilon_{2})g}{1-g^{2}}\sqrt{\frac{\chi+i\gamma_{-}/2}{2\chi}} & \, \frac{-(\epsilon_{1} g^{2}+\epsilon_{2})}{2\chi}\sqrt{\frac{1}{1-g^{2}}} & \, \frac{-(\epsilon_{1} g^{2}+\epsilon_{2})}{2\chi(\chi-i\gamma_{-}/2)}  
	\end{array}\right). \label{Seq.perturb_matrix}
\end{aligned}
\end{equation}

The perturbed eigenvalues are given by the diagonal terms in Eq. \ref{Seq.perturb_matrix}:
\begin{eqnarray}
    \lambda_{0}^{'} &=& \lambda_{0}+\langle L_{0}|H_{p}|R_{0}\rangle=\frac{\epsilon_{1}+\epsilon_{2} g^{2}}{1-g^{2}},\\
    \lambda_{\pm}^{'} &=& \lambda_{\pm}+\langle L_{\pm}|H_{p}|R_{\pm}\rangle = -\frac{i\gamma_{+}}{2}\pm\chi-\frac{\epsilon_{1} g^{2}+\epsilon_{2}}{2\chi(\chi\mp i\gamma_{-}/2)}.
\end{eqnarray}
The corresponding perturbed right eigenvectors are given by:
\begin{flalign}
&\quad \begin{aligned}
        &|R_{0}^{'}\rangle = |R_{0}\rangle + \sum_{s=+,-}\frac{\langle L_{s}|H_{p}|R_{0}\rangle}{\lambda_{0}-\lambda_{s}}|R_{s}\rangle
        = \frac{1}{(1-g^{2})^{5/2}}\left(\begin{array}{c}
                    (1-g^{2})^{2}+i(\epsilon_{1}+\epsilon_{2})\gamma_{-} g^{2} \\
                    -g(1-g^{2})^{2}-i(\epsilon_{1}+\epsilon_{2})\gamma_{-} g   \\
                    (\epsilon_{1}+\epsilon_{2})g(1-g^{2})
                    \end{array} \right), 
\end{aligned}&
\end{flalign}
\begin{flalign}
&\quad\begin{aligned}
        &|R_{\pm}^{'}\rangle = |R_{\pm}\rangle + \sum_{s=0,\mp}\frac{\langle L_{s}|H_{p}|R_{\pm}\rangle}{\lambda_{\pm}-\lambda_{s}}|R_{s}\rangle \\
        &= \sqrt{\frac{1}{\pm2\chi\lambda_{\pm}}}\frac{1}{4\chi^{2}(1-g^{2})^{2}}\left(\begin{array}{c}
                    4\chi^{2}g(1-g^{2})^{2}-4\chi^{2}(\epsilon_{1}+\epsilon_{2})g\lambda_{\mp}-(\epsilon_{1} g^{2}+\epsilon_{2})g(1-g^{2})\lambda_{\pm} \\
                    -4\chi^{2}(1-g^{2})^{2}+4\chi^{2}g^{2}(\epsilon_{1}+\epsilon_{2})\lambda_{\mp}+(\epsilon_{1} g^{2}+\epsilon_{2})(1-g^{2})\lambda_{\pm} \\
                    4\chi^{2}(1-g^{2})^{2}\lambda_{\pm}+(\epsilon_{1} g^{2}+\epsilon_{2})(1-g^{2})^{2}
                    \end{array} \right),   \label{Seq.R_perturbed}
    \end{aligned}&
\end{flalign}
and the perturbed left eigenvectors are given by:
\begin{flalign}
&\quad\begin{aligned}
        &|L_{0}^{'}\rangle = |L_{0}\rangle + \sum_{s=+,-}\frac{\langle R_{s}|H_{p}|L_{0}\rangle}{\lambda_{0}^{*}-\lambda_{s}^{*}}|L_{s}\rangle
        = \frac{1}{(1-g^{2})^{5/2}}\left(\begin{array}{c}
                    (1-g^{2})^{2}-i(\epsilon_{1}+\epsilon_{2})\gamma_{-} g^{2} \\
                    g(1-g^{2})^{2}-i(\epsilon_{1}+\epsilon_{2})\gamma_{-} g \\
                    -(\epsilon_{1}+\epsilon_{2})g(1-g^{2})
                    \end{array} \right), 
\end{aligned}&
\end{flalign}
\begin{flalign}
&\quad\begin{aligned}
        &|L_{\pm}^{'}\rangle = |L_{\pm}\rangle + \sum_{s=0,\mp}\frac{\langle R_{s}|H_{p}|L_{\pm}\rangle}{\lambda_{\pm}^{*}-\lambda_{s}^{*}}|L_{s}\rangle \\
        &= \sqrt{\frac{1}{\pm2\chi\lambda_{\pm}^{*}}}\frac{1}{4\chi^{2}(1-g^{2})^{2}}\left(\begin{array}{c}
                    -4\chi^{2}g(1-g^{2})^{2}+4\chi^{2}(\epsilon_{1}+\epsilon_{2})g\lambda_{\mp}^{*}+(\epsilon_{1} g^{2}+\epsilon_{2})g(1-g^{2})\lambda_{\pm}^{*} \\
                    -4\chi^{2}(1-g^{2})^{2}+4\chi^{2}g^{2}(\epsilon_{1}+\epsilon_{2})\lambda_{\mp}^{*}+(\epsilon_{1} g^{2}+\epsilon_{2})(1-g^{2})\lambda_{\pm}^{*} \\
                    4\chi^{2}(1-g^{2})^{2}\lambda_{\pm}^{*}+(\epsilon_{1} g^{2}+\epsilon_{2})(1-g^{2})^{2}
                    \end{array} \right). \label{Seq.L_perturbed}
    \end{aligned}&
\end{flalign}

For $\gamma,\Gamma \ll 1$, we have
\begin{eqnarray}
    &\langle L_{0}^{'}|R_{\pm}^{'}\rangle=\langle L_{\pm}^{'}|R_{0}^{'}\rangle = \langle L_{+}^{'}|R_{-}^{'}\rangle=\langle L_{-}^{'}|R_{+}^{'}\rangle =\frac{\mathcal{O}(\epsilon_{1}^{2},\epsilon_{2}^{2})}{\chi^{6}},\label{Seq.orth}\\
     &\langle L_{0,\pm}^{'}|R_{0,\pm}^{'}\rangle = 1+\frac{\mathcal{O}(\epsilon_{1}^{2},\epsilon_{2}^{2})}{\chi^{6}},\label{Seq.norm} \\
    &\sum_{s=0,\pm} |R_{s}^{'}\rangle\langle L_{s}^{'}| = \mathbf{I}+\frac{\mathcal{O}(\epsilon_{1}^{2},\epsilon_{2}^{2})}{\chi^{6}}. \label{Seq.com}
\end{eqnarray}
Hence, these perturbed eigenvectors still satisfy the orthogonality, normalization, and completeness conditions.

\subsection{C. State evolution}
\begin{figure}[ht]
	\centering
	\includegraphics[scale=0.6]{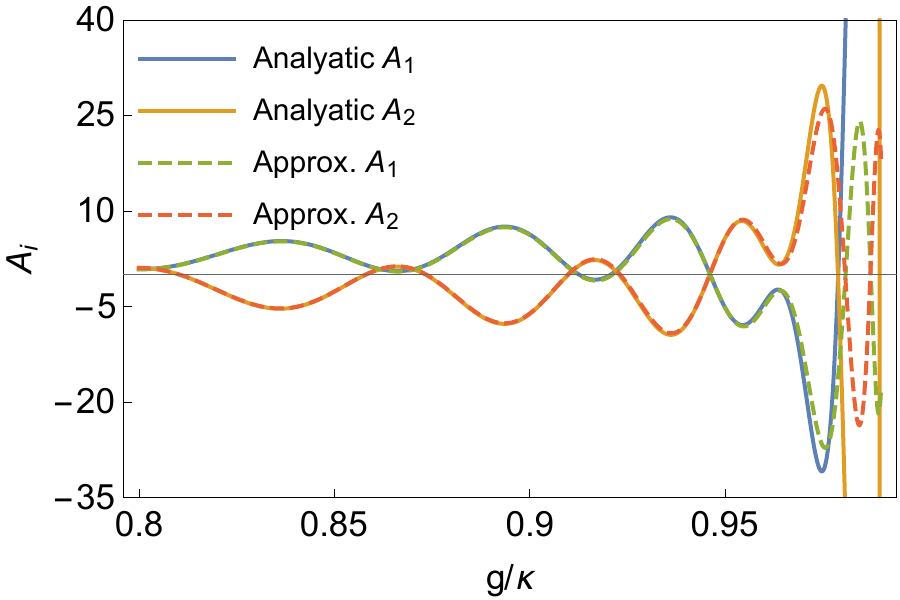}
	\caption{Propargating coefficients $A_{1}$ and $A_{2}$ versus $g$ for fixed time $t=20\pi$. The solid lines shows the numerical simulations of analytic results (Eq. \ref{Seq.propargating_A1}-\ref{Seq.propargating_D}), and the dashed lines represent the results from first-order perturbation theory (Eq. \ref{Seq.A1approx}-\ref{Seq.Dapprox}). In this plot, we set the perturbation strength $\epsilon_{1} = 10^{-3}$ and $\epsilon_{2}=1.5\times10^{-3}$.}
	\label{Sfigure_a1a2}
\end{figure}
Similar to equations \ref{Seq.evolution} and \ref{Seq.eigenvector}, we define the perturbed system evolution matrix $\mathbf{K}$ based on the perturbed biorthogonal eigenvectors:
\begin{equation}
    \mathbf{K} \approx \sum_{s=0,\pm}e^{-i\lambda_{s}^{'}t}|R_{s}^{'}\rangle\langle L_{s}^{'}|. \label{Seq.evolution_perturbed}
\end{equation}
For the lossless case ($\gamma=\Gamma=0$), we have the matrix elements as the propagation coefficients:
\begin{flalign}
&\begin{aligned}
    A_{1}\approx&\frac{e^{-i(\epsilon_{1}+g^{2}\epsilon_{2}) t/\chi^{2}}}{\chi^{2}}-\frac{g^{2}e^{i(g^{2}\epsilon_{1}+\epsilon_{2}) t/(2\chi^{2})}}{16\chi^{8}}[16\chi^{6}+(g^{2}\epsilon_{1}-4\epsilon_{1}-3\epsilon_{2})^{2}]\text{cos}(\chi t)\\
    &-i\frac{g^{2}e^{i(g^{2}\epsilon_{1}+\epsilon_{2}) t/(2\chi^{2})}}{2\chi^{5}}(g^{2}\epsilon_{1}-4\epsilon_{1}-3\epsilon_{2})\text{sin}(\chi t),
\end{aligned}&
\label{Seq.A1approx}
\end{flalign}
\begin{flalign}
&\begin{aligned}
    A_{a}\approx&-\frac{g^{2}(\epsilon_{1}+\epsilon_{2})^{2}e^{-i(\epsilon_{1}+g^{2}\epsilon_{2}) t/\chi^{2}}}{\chi^{6}}+\frac{e^{i(g^{2}\epsilon_{1}+\epsilon_{2}) t/(2\chi^{2})}}{16\chi^{6}}[16\chi^{6}+(g^{2}\epsilon_{1}+\epsilon_{2})^{2}]\text{cos}(\chi t)\\
    &-i\frac{e^{i(g^{2}\epsilon_{1}+\epsilon_{2}) t/(2\chi^{2})}}{2\chi^{3}}(g^{2}\epsilon_{1}+\epsilon_{2})\text{sin}(\chi t),
\end{aligned}&
\end{flalign}
\begin{flalign}
&\begin{aligned}
    A_{2}\approx&-\frac{g^{2}e^{-i(\epsilon_{1}+g^{2}\epsilon_{2}) t/\chi^{2}}}{\chi^{2}}+\frac{e^{i(g^{2}\epsilon_{1}+\epsilon_{2}) t/(2\chi^{2})}}{16\chi^{8}}[16\chi^{6}+(3g^{2}\epsilon_{1}+4g^{2}\epsilon_{2}-\epsilon_{2})^{2}]\text{cos}(\chi t)\\
    &-i\frac{e^{i(g^{2}\epsilon_{1}+\epsilon_{2}) t/(2\chi^{2})}}{2\chi^{5}}(3g^{2}\epsilon_{1}+4g^{2}\epsilon_{2}-\epsilon_{2})\text{sin}(\chi t),
\end{aligned}&
\end{flalign}
\begin{flalign}
&\begin{aligned}
    B\approx&-i\frac{g(\epsilon_{1}+\epsilon_{2})e^{-i(\epsilon_{1}+g^{2}\epsilon_{2}) t/\chi^{2}}}{\chi^{4}}+\frac{ge^{i(g^{2}\epsilon_{1}+\epsilon_{2}) t/(2\chi^{2})}}{16\chi^{7}}[16\chi^{6}-(g^{2}\epsilon_{1}+\epsilon_{2})(g^{2}\epsilon_{1}-4\epsilon_{1}-3\epsilon_{2})]\text{sin}(\chi t)\\
    &+i\frac{g(\epsilon_{1}+\epsilon_{2})e^{i(g^{2}\epsilon_{1}+\epsilon_{2}) t/(2\chi^{2})}}{\chi^{4}}\text{cos}(\chi t),
\end{aligned}&
\end{flalign}
\begin{flalign}
&\begin{aligned}
    C\approx&\frac{ge^{-i(\epsilon_{1}+g^{2}\epsilon_{2}) t/\chi^{2}}}{\chi^{2}}-\frac{ge^{i(g^{2}\epsilon_{1}+\epsilon_{2}) t/(2\chi^{2})}}{16\chi^{8}}[16\chi^{6}-(3g^{2}\epsilon_{1}+4g^{2}\epsilon_{2}-\epsilon_{2})(g^{2}\epsilon_{1}-4\epsilon_{1}-3\epsilon_{2})]\text{cos}(\chi t)\\
    &+i\frac{ge^{i(g^{2}\epsilon_{1}+\epsilon_{2}) t/(2\chi^{2})}}{2\chi^{5}}(g^{2}\epsilon_{1}+2g^{2}\epsilon_{2}+2\epsilon_{1}+\epsilon_{2})\text{sin}(\chi t),
\end{aligned}&
\end{flalign}
\begin{flalign}
&\begin{aligned}
    D\approx&-i\frac{g^{2}(\epsilon_{1}+\epsilon_{2})e^{-i(\epsilon_{1}+g^{2}\epsilon_{2}) t/\chi^{2}}}{\chi^{4}}+\frac{e^{i(g^{2}\epsilon_{1}+\epsilon_{2}) t/(2\chi^{2})}}{16\chi^{7}}[16\chi^{6}+(g^{2}\epsilon_{1}+\epsilon_{2})(3g^{2}\epsilon_{1}+4g^{2}\epsilon_{2}-\epsilon_{2})]\text{sin}(\chi t)\\
    &+i\frac{g^{2}(\epsilon_{1}+\epsilon_{2})e^{i(g^{2}\epsilon_{1}+\epsilon_{2}) t/(2\chi^{2})}}{\chi^{4}}\text{cos}(\chi t).\label{Seq.Dapprox}
\end{aligned}&
\end{flalign}

As shown in Fig. \ref{Sfigure_a1a2}, as long as $\epsilon_{1},\epsilon_{2}\ll \chi^{3}$, the results of the first-order approximation (Eq. \ref{Seq.A1approx}-\ref{Seq.Dapprox}) agree well with the analytical results (Eq. \ref{Seq.propargating_A1}-\ref{Seq.propargating_D}).

\subsection{D. The selection of observable $\hat{X}_{1}\pm\hat{X}_{2}$}
\begin{figure}[ht]
	\centering
	\includegraphics[scale=0.45]{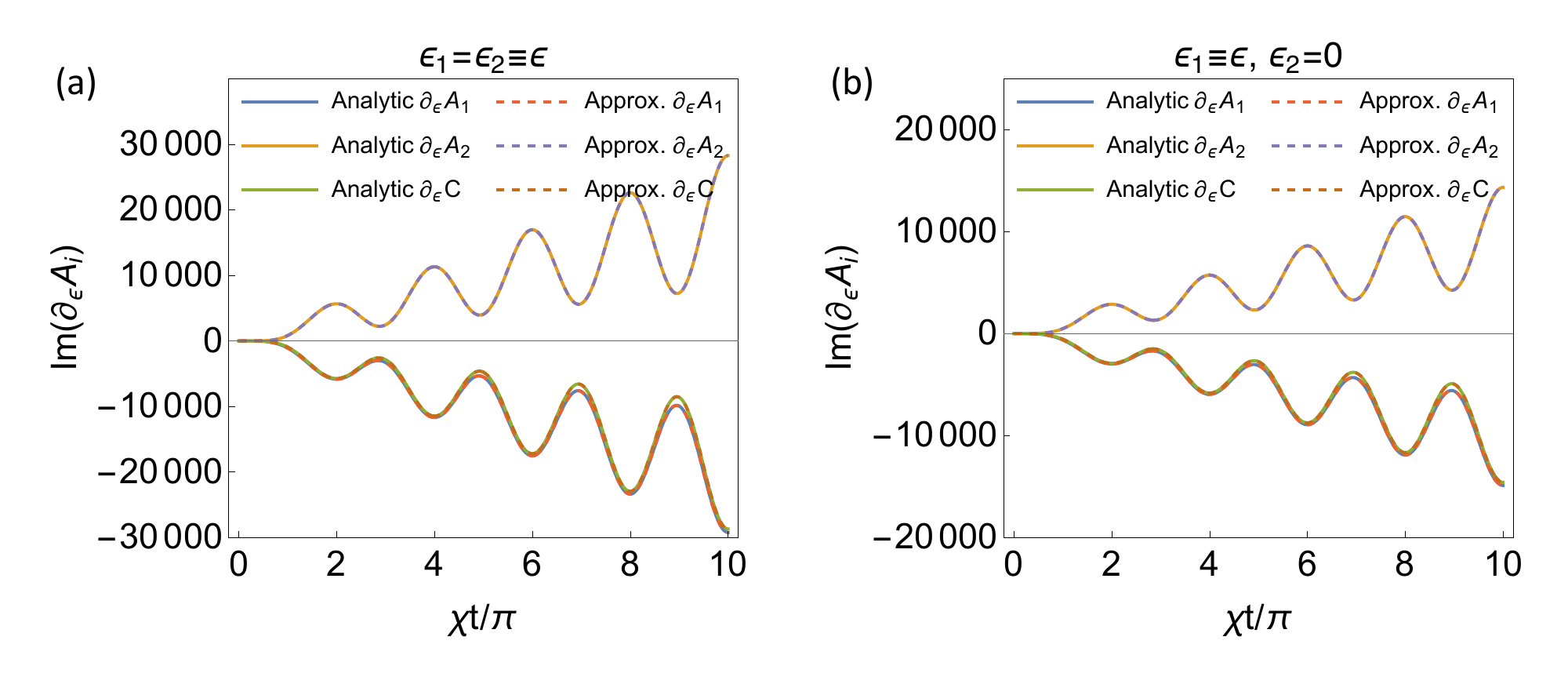}
	\caption{The time evolution of the derivative of propagation coefficents $\{A_{1}, A_{2}, C\}$ with respect to perturbation strength $\epsilon$ for (a) $\epsilon_{1}=\epsilon_{2}\equiv\epsilon$ case and (b) $\epsilon_{1}\equiv\epsilon, \epsilon_{2}=0$ case. The solid and dashed curves represent the results from the analytic solution and first-order perturbation theory, respectively. In both plots, we set $g/\kappa=0.95$ and $\gamma = 0$.}
	\label{Sfigure_partial}
\end{figure}
In the EP3-based sensor, we initially prepare the magnon modes in atomic coherent states. At the working point, we perform the homodyne measurement on the read-out magnon modes. As can be seen from equations 5 and 6 in the main text, the measurement signal is proportional to the derivative of the propagation coefficient (i.e. $\partial_{\epsilon}A_{1}, \partial_{\epsilon}A_{2}$, and $\partial_{\epsilon}C$) which can be calculated from Eq. \ref{Seq.A1approx}-\ref{Seq.Dapprox}.

\textbf{The same perturbations for both atomic ensembles ($\epsilon_{1}=\epsilon_{2}\equiv\epsilon$):} The derivatives of $A_{1}, A_{2}$ and $C$ with respect to $\epsilon$ are given by:
\begin{eqnarray}
    \left.\frac{\partial A_{1}}{\partial\epsilon}\right| _{\epsilon=0} &=& -\frac{i(1+g^{2})\chi t}{\chi^{5}}-\frac{ig^{2}(1+g^{2})\chi t}{2\chi^{5}}\text{cos}(\chi t)-\frac{ig^{2}(g^{2}-7)}{2\chi^{5}}\text{sin}(\chi t), \\
    \left.\frac{\partial A_{2}}{\partial\epsilon}\right| _{\epsilon=0} &=& \frac{ig^{2}(1+g^{2})\chi t}{\chi^{5}}+\frac{i(1+g^{2})\chi t}{2\chi^{5}}\text{cos}(\chi t)+\frac{i(1-7g^{2})}{2\chi^{5}}\text{sin}(\chi t),\\
    \left.\frac{\partial C}{\partial\epsilon}\right| _{\epsilon=0} &=& -\frac{ig(1+g^{2})\chi t}{\chi^{5}}-\frac{ig(1+g^{2})\chi t}{2\chi^{5}}\text{cos}(\chi t) +\frac{i3g(1+g^{2})}{2\chi^{5}}\text{sin}(\chi t). \label{Seq.partial_distributed}
\end{eqnarray}

As shown in Fig. \ref{Sfigure_partial} (a), the derivative of $A_{2}$ has a different sign compared to the derivatives of $A_{1}$ and $C$. Therefore, for the initial states where $\alpha_1=-\alpha_2\equiv i\alpha$ with real $\alpha$, we choose $\hat{X}_1 - \hat{X}_2$ as the observable to maximize the susceptibility:
\begin{equation} \mathcal{S}=\partial_{\epsilon}|\langle\hat{X}_1 - \hat{X}_2\rangle|=\sqrt{2}\alpha|\partial_{\epsilon}\text{Im}(A_{1}-A_{2}+2C)|\sim\chi^{-5}, \label{Seq.max_signal}
\end{equation}
which scales at $\sim\chi^{-5}$ as given explicitly in Eq. 7 of the main text. The properties of noise $\mathcal{N}^{2}=\langle\delta^{2}(\hat{X}_{1}-\hat{X}_{2})\rangle$ and the sensitivity $\delta\epsilon$ are elucidated in the main text.

In contrast, if we choose $\hat{X}_{1}+\hat{X}_{2}$ as the observable, the susceptibility becomes 
\begin{equation}
\begin{aligned}
     \mathcal{S}^{'} = &\sqrt{2}\alpha|\partial_{\epsilon}\text{Im}(A_{1}+A_{2})|\\
     = & \sqrt{2}\alpha \left[ \frac{(1+g^{2})\chi t}{\chi^{3}}-\frac{(1+g^{2})\chi t}{2\chi^{3}}\text{cos}(\chi t)+\frac{(1+g^{2})}{2\chi^{3}}\text{sin}(\chi t)\right]\sim\chi^{-3},
\end{aligned}
\end{equation}
where $A_{1}$ and $A_{2}$ largely cancel with each other. This cancelling reduce the susceptibility $\mathcal{S}^{'}$ to only scales at $\sim \chi^{-3}$. 
Nevertheless, the noise (variance) in this case
\begin{equation}
    \mathcal{N}^{'2}=\langle\delta^{2}(\hat{X}_{1}+\hat{X}_{2})\rangle = \frac{[1+g\text{cos}(\chi t)]^{2}}{(1+g)^{2}}
\end{equation}
exhibits the squeezing property which reaches its maximum squeezing $\mathcal{N}^{'2}_{\text{min}}\sim \chi^{4}$ when $\text{cos}(\chi t)=-1$.
The optimal sensitivity $\delta\epsilon_{\text{opt}}^{'}=\mathcal{N}^{'}_{\text{min}}/\mathcal{S}_{\text{max}}^{'}\sim \chi^{5}$, which also scales at the same order as the Heisenberg limit, is reached for evolution time $\chi t=(2q+1)\pi$.
This sensitivity enhancement over EP2-sensor in this case is attributed to the quantum entanglement between the two atomic ensembles, rather than arising from higher-order EPs.


\textbf{Different perturbations for each atomic ensemble ($\epsilon_{1}\equiv\epsilon, \epsilon_{2}=0$)}: In this case, the derivatives of $A_{1}, A_{2}$ and $C$ with respect to $\epsilon$ are given by:
\begin{eqnarray}
    \left.\frac{\partial A_{1}}{\partial\epsilon}\right| _{\epsilon=0} &=& -\frac{i\chi t}{\chi^{5}}-\frac{ig^{4}\chi t}{2\chi^{5}}\text{cos}(\chi t)-\frac{ig^{2}(g^{2}-4)}{2\chi^{5}}\text{sin}(\chi t), \\
    \left.\frac{\partial A_{2}}{\partial\epsilon}\right| _{\epsilon=0} &=& \frac{ig^{2}\chi t}{\chi^{5}}+\frac{ig^{2}\chi t}{2\chi^{5}}\text{cos}(\chi t)-\frac{i3g^{4}}{2\chi^{5}}\text{sin}(\chi t),\\
    \left.\frac{\partial C}{\partial\epsilon}\right| _{\epsilon=0} &=& -\frac{ig\chi t}{\chi^{5}}-\frac{ig^{3}\chi t}{2\chi^{5}}\text{cos}(\chi t) +\frac{ig(2+g^{2})}{2\chi^{5}}\text{sin}(\chi t).
\end{eqnarray}

As shown in Fig. \ref{Sfigure_partial} (b), the results under this mode are very similar to those obtained with the $\epsilon_{1}=\epsilon_{2}$ case, and thus will not be elaborated upon here. 
The difference lies in the smaller susceptibility and sensitivity.
The reason for this inferior performance is that only one atomic ensemble senses the signal of external perturbation in this case.

\section{S4. Magnon read-out process with SU(2)-type magnon-photon interaction}

In this section, we use the EP3-based sensor as an example to demonstrate how magnons are converted into photons through SU(2)-type magnon-photon interaction.
At the working point $\chi t = 2q\pi$, the interaction for sensing is stopped.
We can expand the state of two magnon modes $\hat{b}_{1}$ and $\hat{b}_{2}$ at the working point under the Fock basis:
\begin{equation}
    \hat{\rho}_{b_{1}b_{2}}=\text{Tr}_{a}[\hat{\rho}_{b_{1}b_{2}a}]=\sum_{n_{1},n_{2}}\sum_{m_{1},m_{2}}\rho_{n_{1},n_{2},m_{1},m_{2}}|n_{1}\rangle_{b_1}|n_{2}\rangle_{b_2\, b_1}\langle m_{1}| _{b_2}\langle m_{2}|.
\end{equation}
Here, $\text{Tr}_{a}[\cdot]$ means trace over mode $\hat{a}$ and $\hat{\rho}_{b_{1}b_{2}a}$ is the three-mode magnon-photon hybrid state of the time.

\begin{figure}[ht]
	\centering
	\includegraphics[scale=0.5]{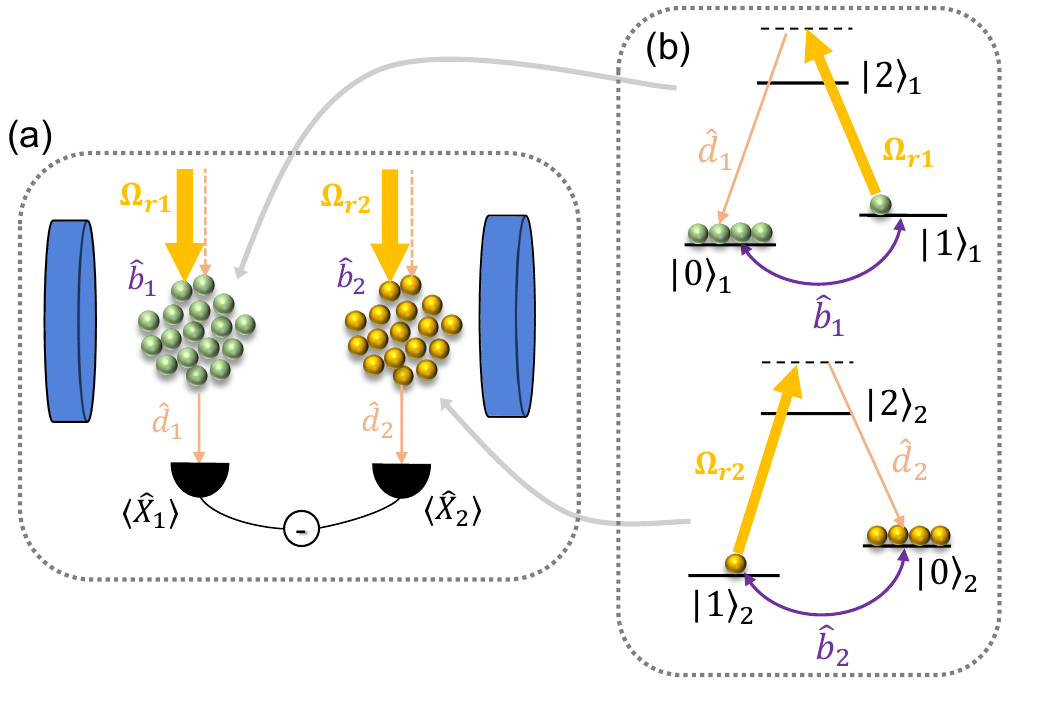}
	\caption{(a) The schematic diagram of magnon read-out process and the homodyne detection on the read-out light mode. (b) The energy level of SU(2)-type Raman interaction for each atomic ensemble.}
	\label{Sfigure_readout}
\end{figure}

The reading process is achieved through the SU(2)-type Raman interaction, whose Hamiltonian is given by:
\begin{equation}
    \hat{\mathcal{H}}_{read} = \sum_{j=1,2}i\theta_{j}(\hat{d}_{j}^{\dagger}\hat{b}_{j}-\hat{b}_{j}^{\dagger}\hat{d}_{j}),
\end{equation} 
where $\hat{d}_{j}$ are the light modes for homodyne detection and initially prepared in vacuum states. 
$\theta_{j} = \sqrt{N_{a}}|g_{bs}\Omega_{rj}|/\Delta$ is the effective coupling coefficient. Here for convenience, we assume $\theta_{1}=\theta_{2}=\theta$.
The evolution operator of this process is therefore $\hat{U}=\text{exp}(-i\hat{\mathcal{H}}_{read}t)$. 
After interaction time $t$, the output state of two magnon and output photon modes is given by:
\begin{equation}
\begin{aligned}
    \hat{\rho}_{out} = & \,\hat{U}\hat{\rho}_{b_{1}b_{2}}\otimes\hat{\rho}_{d_{1}d_{2}}\hat{U}^{\dagger}\\
    =& \sum_{n_{1},n_{2}}\sum_{m_{1},m_{2}}\rho_{n_{1},n_{2},m_{1},m_{2}}\hat{U}|n_{1}\rangle_{b_1}|n_{2}\rangle_{b_2\, b_1}\langle m_{1}| _{b_2}\langle m_{2}|\otimes|0\rangle_{d_{1}}|0\rangle_{d_{2}\,d_{1}}\langle 0|_{d_{2}}\langle 0|\hat{U}^{\dagger}\\
    =& \sum_{n_{1},n_{2}}\sum_{m_{1},m_{2}}\frac{\rho_{n_{1},n_{2},m_{1},m_{2}}}{\sqrt{n_{1}!n_{2}!m_{1}!m_{2}!}}\hat{U}(\hat{b}_{1}^{\dagger})^{n_1}(\hat{b}_{2}^{\dagger})^{n_2}\hat{U}^{\dagger}\hat{U}|0,0,0,0\rangle\langle 0,0,0,0|\hat{U}^{\dagger}\hat{U}(\hat{b}_{1})^{m_1}(\hat{b}_{2})^{m_2}\hat{U}^{\dagger}\\
    =&\sum_{n_{1},n_{2}}\sum_{m_{1},m_{2}}\frac{\rho_{n_{1},n_{2},m_{1},m_{2}}}{\sqrt{n_{1}!n_{2}!m_{1}!m_{2}!}}[\text{cos}(\theta t)\hat{b}_{1}^{\dagger}+\text{sin}(\theta t)\hat{d}_{1}^{\dagger}]^{n_1}[\text{cos}(\theta t)\hat{b}_{2}^{\dagger}+\text{sin}(\theta t)\hat{d}_{2}^{\dagger}]^{n_2}|0,0,0,0\rangle\\
    &\otimes \langle 0,0,0,0|[\text{cos}(\theta t)\hat{b}_{1}+\text{sin}(\theta t)\hat{d}_{1}]^{m_1}[\text{cos}(\theta t)\hat{b}_{2}+\text{sin}(\theta t)\hat{d}_{2}]^{m_2}.
\end{aligned}
\end{equation}
When $\theta t=\pi/2$, the state becomes:
\begin{equation}
\begin{aligned}
     \hat{\rho}_{out} =& \sum_{n_{1},n_{2}}\sum_{m_{1},m_{2}}\rho_{n_{1},n_{2},m_{1},m_{2}}|0\rangle_{b_{1}}|0\rangle_{b_{2}\,b_{1}}\langle 0|_{b_{2}}\langle 0|\otimes|n_{1}\rangle_{d_1}|n_{2}\rangle_{d_2\, d_1}\langle m_{1}| _{d_2}\langle m_{2}|\\
     =& \hat{\rho}_{b_{1}b_{2}}^{'}\otimes\hat{\rho}_{d_{1}d_{2}}^{'}.
\end{aligned}
\end{equation}
Here, $\hat{\rho}_{b_{1}b_{2}}^{'}=\hat{\rho}_{d_{1}d_{2}}=|0,0\rangle\langle 0,0|$ and $\hat{\rho}_{d_{1}d_{2}}^{'}=\hat{\rho}_{b_{1}b_{2}}$, which means both magnon modes are converted into photons and back to their vacuum states. Therefore, the homodyne measurements performed on the read-out light modes $\hat{d}_{i}$ can fully extract the information carried by the atomic ensembles.

\section{S5. Quantum Fisher information and Gaussian decomposition}
To determine the ultimate precision of our proposed sensor and to understand the physical origin of the enhanced sensitivity, we investigate the quantum Fisher information (QFI) and analyze the results from the perspective of Gaussian decomposition.

\subsection{A. Quantum Fisher information} 
We consider the magnon modes $\hat{b}_1$, $\hat{b}_2$ and the cavity mode $\hat{a}$ to be initially prepared in spin coherent states $|i\alpha\rangle_{b_1}$, $|-i\alpha\rangle_{b_2}$ and vacuum state $|0\rangle_{a}$, respectively. In the absence of any loss, the magnon-photon hybrid state remains a pure Gaussian state in the whole process \cite{weedbrook2012gaussian}.
The state can be fully characterized by the first moment (displacement) vector $\langle\Vec{\mu}\rangle=(\langle\hat{X}_{1}\rangle,\langle\hat{P}_{1}\rangle,\langle\hat{X}_{2}\rangle,\langle\hat{P}_{2}\rangle,\langle\hat{X}_{a}\rangle,\langle\hat{P}_{a}\rangle)^{T}$ and the second moment (covariance matrix) $\mathbf{\Lambda}_{ij}=\langle\{\hat{\mu}_{i},\hat{\mu}_{j}\}\rangle/2-\langle\hat{\mu}_{i}\rangle\langle\hat{\mu}_{j}\rangle$.


The sensitivity of parameter $\epsilon$ is bounded by the inverse of the QFI $\mathcal{I}(\epsilon)$ of the state through the quantum Cramér-Rao bound: $\delta^{2}\epsilon \geq 1/\mathcal{I}(\epsilon)$.
For Gaussian process, the QFI takes the form:
\begin{equation}
\begin{aligned}
    \mathcal{I}(\epsilon)&=\mathcal{I}_{\mathbf{\Lambda}}(\epsilon)+\mathcal{I}_{\mu}(\epsilon),\\
    &= \frac{1}{2}\text{Tr}\left[\mathbf{\Phi}_{\epsilon} \frac{d\mathbf{\Lambda}}{d\epsilon}\right]+\left(\frac{d\langle\Vec{\mu}\rangle}{d\epsilon}\right)^{T} \mathbf{\Lambda}^{-1}\frac{d\langle\Vec{\mu}\rangle}{d\epsilon}. \label{Seq.QFI}
\end{aligned}
\end{equation}
Here, $\mathbf{\Phi}_{\epsilon}$ is implicitly determined by $d\mathbf{\Lambda}/d\epsilon=\mathbf{\Lambda}\mathbf{\Phi}_{\epsilon}\mathbf{\Lambda}-\mathbf{\Omega}\mathbf{\Phi}_{\epsilon}\mathbf{\Omega}^{T}$, $\mathbf{\Omega}=\mathbf{I}_{3}\otimes(0,1;-1,0)$ is the fundamental symplectic matrix.

It is obvious that the first term $\mathcal{I}_{\mathbf{\Lambda}}$ in Eq. \ref{Seq.QFI} is determined solely by the change of the covariance matrix. The second term $\mathcal{I}_{\mu}$ is related to the change of the displacement of the state. 
As shown in Fig. \ref{Sfigure_QFI} (a), when \(\alpha\) is sufficiently large, $\mathcal{I}_{\mu}\sim \chi^{-10}$ dominates the QFI, indicating that the change of noise barely contributes to the sensitivity. Therefore, simply measuring the change of displacement is a sub-optimal measurement strategy.

\begin{figure}[htbp]
	\centering
	\includegraphics[scale=0.45]{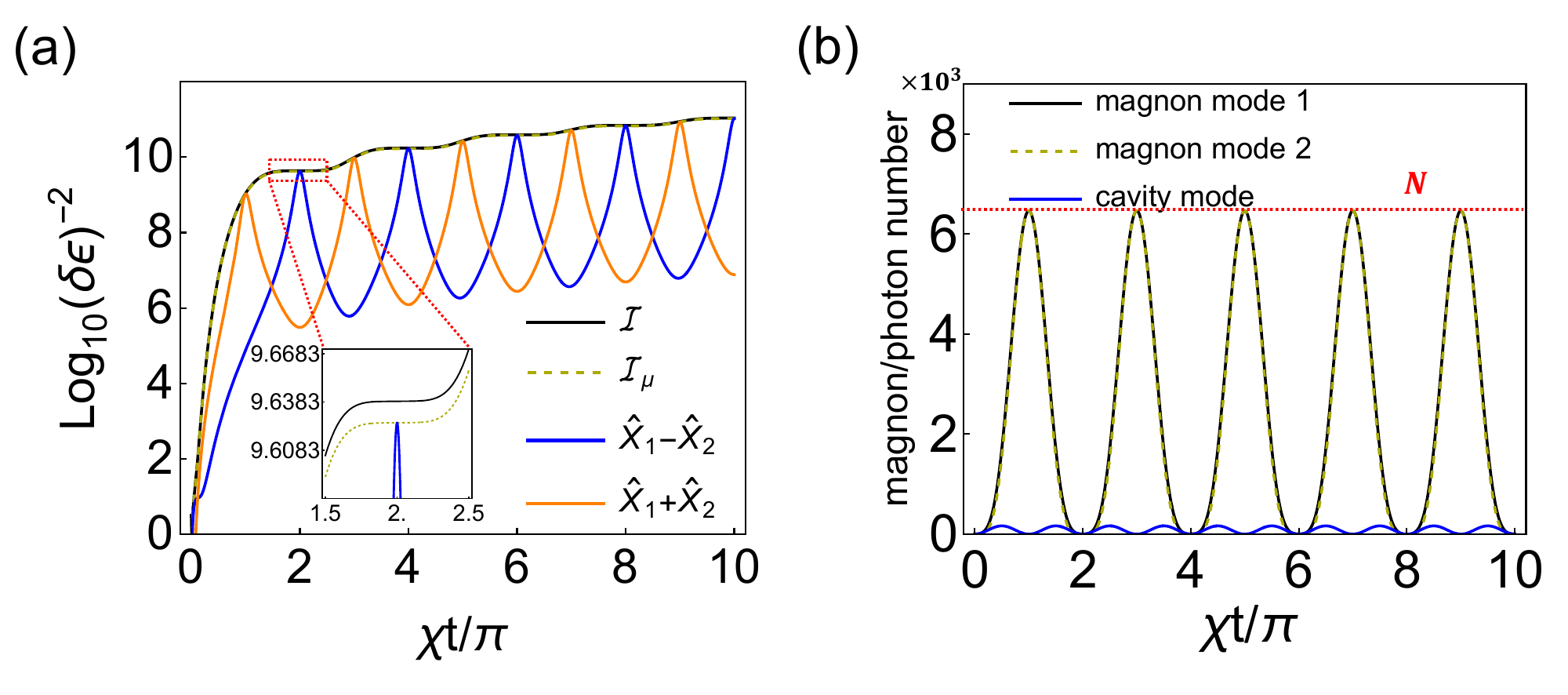}
	\caption{(a) The time evolution of the QFI and the inverse variance $(\delta\epsilon)^{-2}$ of measuring observable $\hat{X}_{1}\pm\hat{X}_{2}$. (b) The time evolution of the magnon number and cavity photon number. The magnon number is bounded by $N=[\alpha^{2}(\kappa+g)^{4}+4\kappa^{2}g^{2}]/\chi^{4}\sim\chi^{-4}$. In both plots, we set $g/\kappa=0.95$, $\kappa=1$ and $\alpha = 2$.}
	\label{Sfigure_QFI}
\end{figure}

In order to study the scaling of QFI in terms of particle number and evolution time, we calculate the magnon and photon number during the evolution. As shown in Fig. \ref{Sfigure_QFI} (b), the magnon number oscillates over time with a period of \( T = 2\pi/\chi \). 
The total magnon number $N_{magnon}=N_1 + N_2$ is given by:
\begin{equation}
\begin{aligned}
    N_{magnon}=&N_1 +N_2\\
    &=\frac{\alpha^{2}(\kappa+g)^{2}\{(\kappa^{2}+g^2)[1+\text{cos}^{2}(\chi t)]-4\kappa g\text{cos}(\chi t)\}+4\kappa^2 g^2[1-\text{cos}(\chi t)]-g^2(\kappa^2+g^2)\text{sin}^{2}(\chi t)}{\chi^4} \\
     &\leq 2\frac{\alpha^{2}(\kappa+g)^{4}+4\kappa^{2}g^{2}}{\chi^4}=2N.
\end{aligned}
\end{equation}
Here, $N_{i}$ represents the magnon number of mode $i$.
The intracavity photon number $N_a$ oscillates over time with a period of \( T = \pi/\chi \) and is given by:
\begin{equation}
    N_{a}=\frac{\alpha^{2}(\kappa+g)^2+g^2}{\chi^2} \text{sin}^{2}(\chi t).
\end{equation}
Here, $N_1-N_2-N_a$ is the a conservative number in our system.
The results show that the QFI ($\sim \chi^{-10}$) scales at the same order as the Heisenberg limit $(2N+N_a)^{2}t^{2}\sim\chi^{-10}$.
The total number of magnons and photons determines the rate of increase in QFI. 
For $t=q\pi/\chi$, the intracavity photon number is zero, therefore only the magnon number contributes to the QFI growth.
At \( t = 2q\pi/\chi \), the total magnon number reaches its minimum, causing the QFI to increase slowly. In contrast, at \( t = (2q+1)\pi/\chi \), the total magnon number reaches its maximum, resulting in a rapid increase in the QFI.

For EP-based strategy, i.e., measuring $\hat{X}_{1}-\hat{X}_{2}$ at $t=2q\pi/\chi$, the covariance matrix is $\mathbf{\Lambda}_{\epsilon=0}=\mathbf{I}/2$. Therefore, only the change of the displacement $d\langle\Vec{\mu}\rangle/d\epsilon \sim \chi^{-5}$ contributes to $\mathcal{I}_{\mu}$. 
In contrast, for squeezing-based strategy, i.e., measuring $\hat{X}_{1}+\hat{X}_{2}$ at $t=(2q+1)\pi/\chi$, the change of the measurement outcome only scales at $\chi^{-3} \sim \sqrt{N}t$ which is unamplified.
The enhanced sensitivity in this strategy relies on the squeezed noise $\langle\delta^{2}(\hat{X}_{1}+\hat{X}_{2})\rangle\sim \chi^4$ through the correlation in quantum fluctuation between atoms.

\subsection{B. Gaussian decomposition} 
For the system with EPn, the state can be realized by preparing $n$ vacuum states, applying multimode interferometers and one-mode squeezers and finally displacing them by $\Bar{\mathbf{x}}$, as described by the following equation \cite{weedbrook2012gaussian}:
\begin{equation}
    |\psi\rangle = \hat{D}(\Bar{\mathbf{x}})\,\hat{U}_{K}\,\left[\bigotimes_{k=1}^{n}\hat{S}(r_k)\right]\,\hat{U}_{L}\,|0\rangle^{\otimes n}. \label{Seq.Gaussian_decomposition}
\end{equation}
Here, $\hat{D}(\Bar{\mathbf{x}})$ is the displacement operator with $\Bar{\mathbf{x}} \in \Bbb{R}^{2n}$, $\hat{U}_{K}$ and $\hat{U}_{L}$ are the passive canonical unitaries that describe multiport interferometers (e.g., the beam splitter in the case of two modes), and $\bigotimes_{k=1}^{n}\hat{S}(r_k)$ is a parallel
set of $n$ one-mode squeezers.
In terms of quadrature operators, this Gaussian unitary transformation is more simply described by an affine map: $\Vec{\mu}\rightarrow \mathbf{\Sigma}\,\Vec{\mu}+\Bar{\mathbf{x}}$, where $\mathbf{\Sigma}$ is $2n\times 2n$ real symplectic matrix.

For a general EPn system, solving Gaussian decomposition explicitly is quite challenging. 
\textit{Here, we conduct Gaussian decomposition for a two-mode system with an EP2 to identify which part of Eq. \ref{Seq.Gaussian_decomposition} contributes to the enhanced sensitivity of the EP-based sensor. 
For higher-order n-mode system, the principle is the same.}
This two-mode system consists of one magnon mode and one cavity mode, which interact with each other via SU(1,1)-type interaction. The interaction Hamiltonian of this system can be the same as in \cite{luo2022quantum} after some gauge transformations:
\begin{equation}
    \hat{\mathcal{H}}_{EP2}=(\delta+\epsilon)(\hat{a}^{\dagger}\hat{a}+\hat{b}^{\dagger}\hat{b}) + ig(\hat{a}^{\dagger}\hat{b}^{\dagger}-\hat{a}\hat{b}).
\end{equation}
Here, \(\delta\) is the tunable detuning, and \(\epsilon\) is the external perturbation to be measured.
In the absence of external perturbation $\epsilon=0$, EP2 appears when $\delta=g$. 
Near this EP2, we can estimate the small perturbation $\epsilon$ by measuring the changes in the system.

For initial state $|\alpha\rangle_{b}, \alpha\in\Bbb{R}$, the displacement $\Bar{\mathbf{x}}$ is given by:
\begin{equation}
    \Bar{\mathbf{x}} =\sqrt{2}\alpha \left[\text{Re}(B), \text{Im}(B), \text{Re}(A),  \text{Im}(A)\right]^T.
\end{equation}
The symplectic matrix $\mathbf{\Sigma}$ can be further decomposed using the Bloch–Messiah/Euler decomposition $\mathbf{\Sigma}=\mathbf{K}\left[\bigoplus_{k=1}^{n}\mathbf{S}(r_k)\right]\,\mathbf{L}$ \cite{braunstein2005squeezing}:
\begin{equation}
    \mathbf{K}=\frac{1}{\sqrt{2}}\left(\begin{array}{cc}
        \mathbf{I}_2 & -\mathbf{I}_2  \\
         \mathbf{R}(\phi) & \mathbf{R}(\phi)
    \end{array}\right), \, \bigoplus_{k=1}^{n}\mathbf{S}(r_k)=\left(\begin{array}{cc}
        \mathbf{S}(-r) & 0 \\
         0 & \mathbf{S}(r)
    \end{array}\right), \,\mathbf{L}=\frac{1}{\sqrt{2}}\left(\begin{array}{cc}
        \mathbf{R}(\phi) & \mathbf{I}_2 \\
         -\mathbf{R}(\phi) & \mathbf{I}_2
    \end{array}\right)
\end{equation}
$\mathbf{K}$ and $\mathbf{L}$ are symplectic and orthogonal, while $\mathbf{S}(r_k)$ is a set of one-mode squeezing matrices.
Here, $\phi=\text{arctan}[\text{Im}(A)/\text{Re}(A)]$, $\mathbf{R}(\phi)=[\text{cos}(\phi),\text{sin}(\phi);-\text{sin}(\phi),\text{cos}(\phi)]$, $\mathbf{S}(r)=\text{diag}\{e^{-r},e^{r}\}$ and $r=\mp\text{In}(|A|\pm |B|)$.
$A$ and $B$ can be expressed by:
\begin{eqnarray}
    A(\epsilon, t)&=&\text{cos}(\chi t)-i\frac{\delta+\epsilon}{\chi}\text{sin}(\chi t),\\
    B(\epsilon, t)&=&\frac{g}{\chi}\text{sin}(\chi t),\\
    \chi(\epsilon) &=& \sqrt{(\delta+\epsilon)^{2}-g^{2}}.
\end{eqnarray}

Near EP2 ($\chi\rightarrow 0$) and at the working point $\chi t=2q\pi$, since $\text{Re}(\partial_{\epsilon}B)=2\delta gq\pi/\chi^3$ and $\text{Im}(\partial_{\epsilon}A)=2\delta^{2} q\pi/\chi^3$, the displacement $\Bar{\mathbf{x}}$ changes significantly under the perturbation $\epsilon$, i.e. 
\begin{equation}
    \partial_{\epsilon} \,\Bar{\mathbf{x}}\sim \alpha\chi^{-3}.
\end{equation}
This significant change in displacement can be considered as the amplification of the sensor's signal by the EP. 
While the squeezing factor $r(t=2q\pi/\chi)=0$, which means that the noise is at the vacuum noise level.

Notice that, the variation of squeezing factor in presence of perturbation $\partial_{\epsilon}\, r\sim \chi^{-3}$, which contributes to $\mathcal{I}_\Lambda$, is also large.
However, for sufficiently large $\alpha$, we can conclude that the change in the system's squeezing (i.e., the variation of the covariance matrix) barely contributes to the overall sensitivity by comparing $\mathcal{I}$ (QFI) and $\mathcal{I}_\mu$ in Fig. \ref{Sfigure_QFI} (a).
Therefore, the primary contribution to the enhanced sensitivity in EP-based sensor comes from the significant change in displacement within the phase space.
Accordingly, in the main text, we selected a suitable observable that maximally utilizes the change in displacement, while the variation of measurement noise remains negligible.

For a general EPn system, the symplectic matrix always evolves to the identity matrix $\mathbf{\Sigma}=\mathbf{I}_{2n}$, meaning that the squeezing factor for each mode is zero.
While the EPn significantly amplifies the change in displacement with a general scaling law $\sim\chi^{1-2n}$.
Therefore, the physical mechanism behind it is the same as in the EP2 case.
In summary, from the perspective of Gaussian decomposition, the enhanced sensitivity in EP-based sensors primarily arises from the displacement operator $\hat{D}(\Bar{\mathbf{x}})$ rather than the squeezing operator $\hat{S}(r)$.

\section{S6. The impacts of losses}
In this section, we consider the impacts of losses on the system, including both internal and external losses. Internal losses consist of cavity leakage and magnon decay, which affect the evolution of the system. External losses, on the other hand, do not influence the system’s evolution but arise from imperfections in the detection process.

\subsection{A. Internal losses}
Here, we investigate the influence of internal losses on the sensitivity, which is described by the following Heisenberg-Langevin equations:
\begin{equation}
    \partial_{t}\left(\begin{array}{c}
                    \hat{b}_{1}(t) \\
                     \hat{b}_{2}^{\dagger}(t) \\
                     \hat{a}^{\dagger}(t)
                    \end{array} \right) =-i h_{D}^{(3,1)} \left(\begin{array}{c}
                    \hat{b}_{1}(t) \\
                     \hat{b}_{2}^{\dagger}(t) \\
                     \hat{a}^{\dagger}(t)
                    \end{array} \right) + \left(\begin{array}{c}
                    \sqrt{2\Gamma}\hat{f}_{1}(t) \\
                     \sqrt{2\Gamma}\hat{f}^{\dagger}_{2}(t) \\
                     \sqrt{2\gamma}\hat{f}^{\dagger}_{a}(t)
                    \end{array} \right), \label{Seq.langevin_eq}
\end{equation}
with $\gamma$ and $\Gamma$ represent the decay rates of the cavity mode and the magnon modes, respectively.
Here, $\hat{f}_{k}(t)$ are Langevin noises for each mode satisfying $\langle\hat{f}_{k}(t)\hat{f}^{\dagger}_{k}(t')\rangle=\delta(t-t'), k=1,2,a$.
The solution of Eq. \ref{Seq.langevin_eq} can be obtained with direct integration:
\begin{equation}
    \left(\begin{array}{c}
                    \hat{b}_{1}(t) \\
                     \hat{b}_{2}^{\dagger}(t) \\
                     \hat{a}^{\dagger}(t)
                    \end{array} \right) = \mathbf{K}(t)\left(\begin{array}{c}
                    \hat{b}_{1}(0) \\
                     \hat{b}_{2}^{\dagger}(0) \\
                     \hat{a}^{\dagger}(0)
                    \end{array} \right) + \int_{0}^{t} \mathbf{K}(t-\tau)\left(\begin{array}{c}
                    \sqrt{2\Gamma}\hat{f}_{1}(\tau) \\
                     \sqrt{2\Gamma}\hat{f}^{\dagger}_{2}(\tau) \\
                     \sqrt{2\gamma}\hat{f}^{\dagger}_{a}(\tau)
                    \end{array} \right) d\tau.
\end{equation}
We solve the above Heisenberg-Langevin equations numerically and obtain the sensitivity under the influence of internal losses. The results show that the internal losses cause the system to undergo damped oscillations in both susceptibility and noise.
The SU(1,1)-type Raman interaction in the system amplifies the Langevin noises, leading to a constant noise that degrades the sensitivity.

\begin{figure}[b]
	\centering
	\includegraphics[scale=0.4]{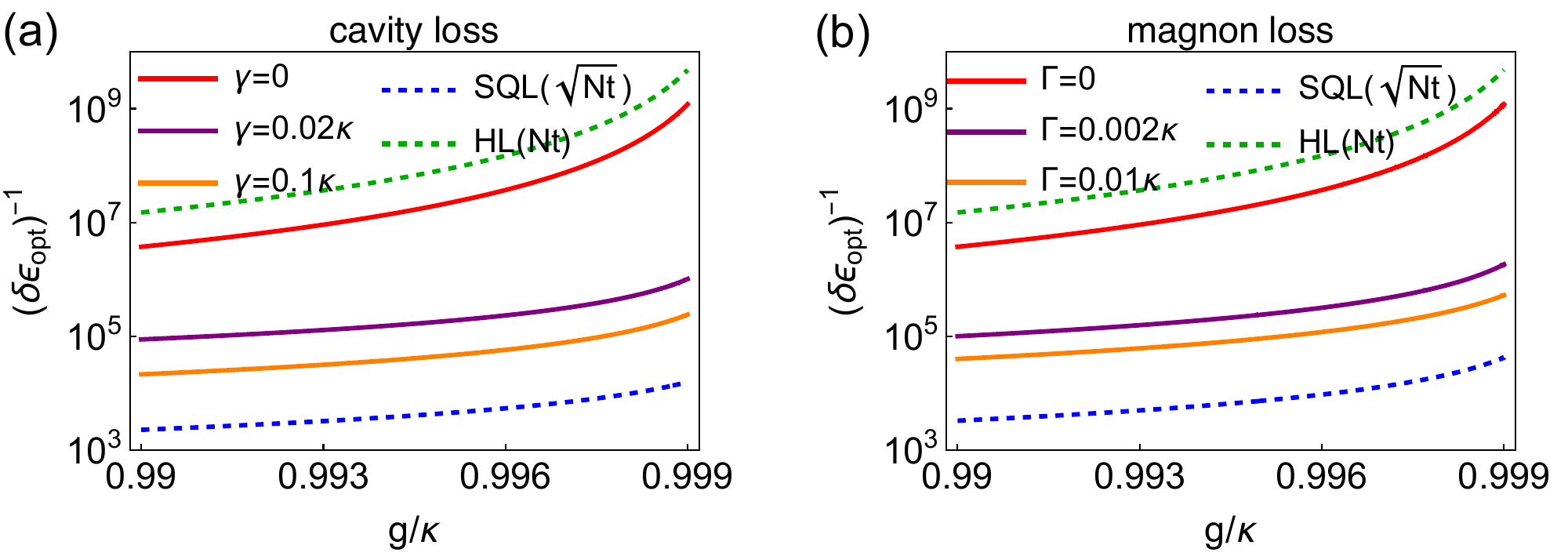}
	\caption{The reciprocal of optimal sensitivity $(\delta\epsilon_{\text{opt}})^{-1}$ as a function of $g$ under the influence of (a) cavity loss and (b) magnon loss. Here, we set $t=2\pi/\chi$, $\kappa=1$ and $\alpha = 2$.}
	\label{Sfigure_loss}
\end{figure}


In Fig. \ref{Sfigure_loss}, we plot the optimal sensitivity varying with $g$ at the first working point $\chi t=2\pi$.
The results show that both cavity loss and magnon loss reduce the system's optimal sensitivity and alter the sensitivity's scaling factor as they destroy the Hermicity of the system.
The difference is that magnon losses have a greater impact on the system's performance because the magnon number is much higher than the intracavity photon number, resulting in a higher error rate.
Under the influence of internal losses, the scaling factor of optimal sensitivity gradually approaches the scaling of the standard quantum limit (SQL) $1/\sqrt{Nt}$.
However, quantum-enhanced measurement can still be achieved at experimentally reasonable decay rates, i.e., $\gamma=0.1\kappa\sim 2\pi\times50\text{kHz}$ and $\Gamma=0.01\kappa\sim2\pi\times5\text{kHz}$.

\begin{figure}[t]
	\centering
	\includegraphics[scale=0.4]{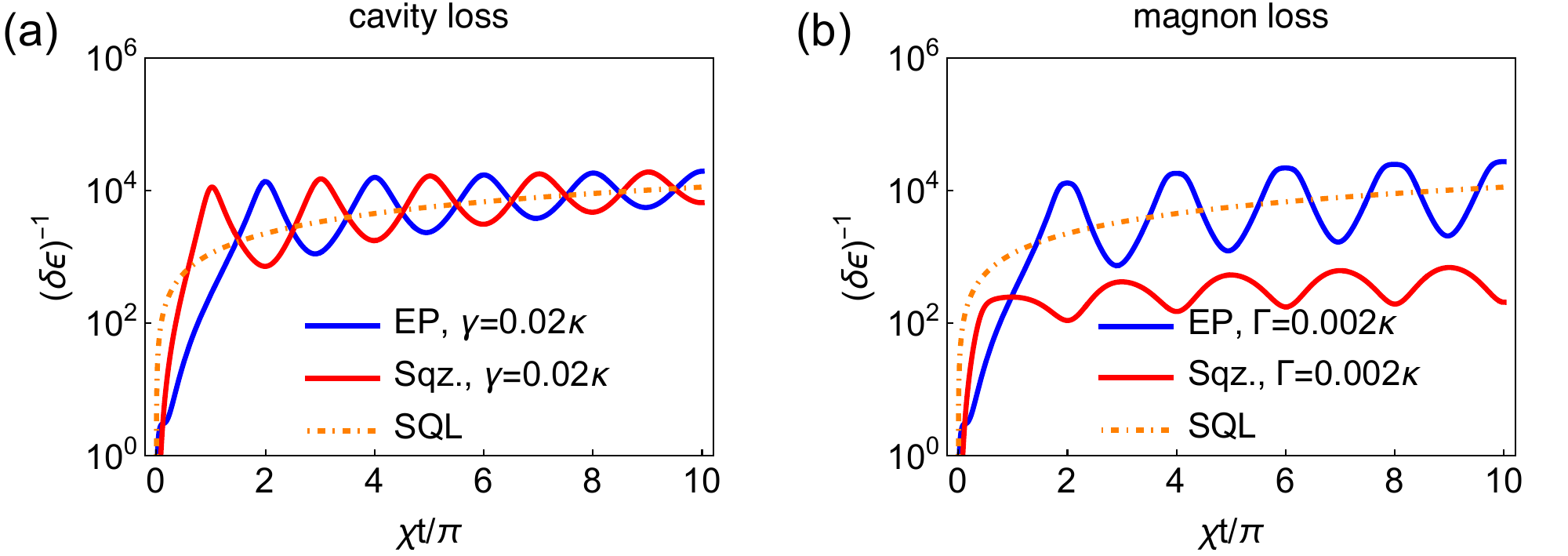}
	\caption{The time evolution of the reciprocal of sensitivity $(\delta\epsilon)^{-1}$ of measuring observable $\hat{X}_{1}\pm\hat{X}_{2}$ under the influence of (a) cavity loss and (b) magnon loss. Here, we set $g=\kappa=0.95$, $\kappa=1$ and $\alpha = 2$.}
	\label{Sfigure_internal_compare}
\end{figure}

We further compare the performance of the EP-based (observable $\hat{X}_{1}-\hat{X}_{2}$) and squeezing-based strategies (observable $\hat{X}_{1}+\hat{X}_{2}$) under internal losses.
The impact of cavity loss is the same for both strategies [see Fig. \ref{Sfigure_internal_compare} (a)], whereas magnon loss affects the EP-based strategy much less than the squeezing-based one [see Fig. \ref{Sfigure_internal_compare} (b)].
In the EP-based strategy, the EP-induced signal amplification still remains even with the extra Langevin noise induced by cavity and magnon decay, making this strategy hold loss tolerance.
In contrast, for the squeezing-based strategy, the Langevin noises induced by magnon decay disrupts the fragile quantum entanglement between atomic ensembles, leading to substantial performance degradation.

\begin{figure}[b]
	\centering
	\includegraphics[scale=0.4]{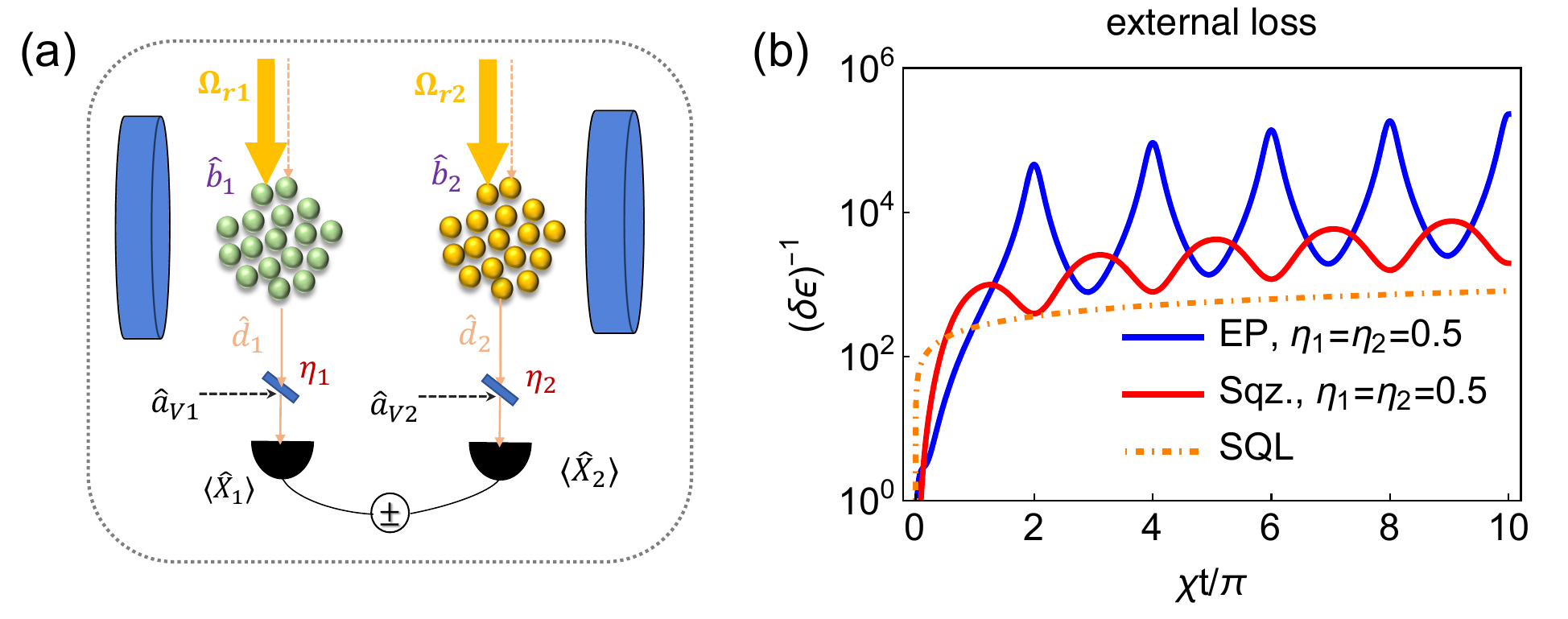}
	\caption{(a) The schematic diagram of the external losses induced in magnon readout processes and detector's imperfections. (b) The influence of external losses. Here, we set $g/\kappa=0.95$ and $\alpha = 2$.}
	\label{Sfigure_external}
\end{figure}

\subsection{B. External losses}
Here, we investigate the influence of external losses on the sensitivity, such as those arising from the magnon readout processes and the imperfections of the detectors.
We can model the external loss of mode $i$ as a beam splitter with transmissivity $\eta_i$ that mixes the magnon field $i$ with the vacuum noise, as depicted in Fig. \ref{Sfigure_external} (a).

In the absence of any losses, the sensitivity of EP-based strategy and that of squeezing-based one both scale at the same order as the Heisenberg limit.
However, when external losses are considered, EP-based strategy exhibits an advantage in loss tolerance compared to squeezing-based one, as shown in Fig. \ref{Sfigure_external} (b). EP sensors maintain the same sensitivity scaling under high external losses. This is because, in the EP scheme, the sensitivity enhancement relies solely on the amplification of the displacement signal, with the noise at the working point being at the same level as the vacuum noise. The external loss linearly attenuates the displacement signal (by a constant factor $\eta_i$), so it does not alter the scaling of the sensitivity. In contrast, the sensitivity enhancement for the squeezing scheme depends on noise squeezing induced by quantum correlation. The loss-induced-mixing of the squeezed field with vacuum noise reduces the squeezing factor to $r'$ nonlinearly ($e^{-2r'} = \eta e^{-2r}+1-\eta$), leading to substantial performance deterioration in squeezing-based sensors.

\section{S7. The scaling of sensitivity $\delta\epsilon$ near $n$-th order EP}
In this section, we briefly analyze the scaling factor of the sensitivity $\delta\epsilon$ with the perturbation $\epsilon$ (and the corresponding $\chi$) for our proposed EPn-based atom-cavity sensor. 
Similar to the EP3-based sensor, we can obtain perturbation information by performing homodyne detection on the read-out magnon modes. 
We consider coherent initial state for the magnon modes. The susceptibility is then given by the derivative of the propagation coefficients with respect to \(\epsilon\): 
\begin{equation}
    \mathcal{S} \propto \left|\frac{\partial \mathbf{K}_{jk}}{\partial\epsilon}\right|,
\end{equation}
where $\mathbf{K}_{jk}$ is given by Eq. \ref{Seq.Kelements_jj}-\ref{Seq.Kelements_jk}.

When the system is at EPn ($\lambda_{1}=\cdots=\lambda_{n}=\lambda_{0}$) and exhibits an $n$-th order response to the perturbation $\epsilon$, the splitting corresponding to each eigenvalue is:
\begin{equation}
    \Delta\lambda_{p} = \lambda_{p} - \lambda_{0} \propto \text{exp}(i\frac{2\pi}{n}p)\,\epsilon^{\frac{1}{n}}
\end{equation}
with $p = 1, \cdots, n$.
The denominator of $\mathbf{K}_{jk}$ is then $\prod_{q\neq p}^{n}(\lambda_{p}-\lambda_{q}) \propto \epsilon^{\frac{n-1}{n}}$.
We note that when $|\lambda_{0}| \gg |\epsilon| \neq 0$, the polynomial $F_{n}^{(jk)}$ can be regarded as a quantity independent of $\epsilon$. When $\lambda_{0} = 0$, since $|a_{0}| \gg |\epsilon| \neq 0$, the polynomial $F_{n}^{(jk)}$ remains a quantity independent of $\epsilon$.
We therefore assert that $\mathbf{K}_{jk}$ scales at $\sim \epsilon^{-(n-1)/n}$ and the susceptibility is $\mathcal{S} \propto \epsilon^{-2+1/n}$.

As we approach the system's EPn from the stable side (where all eigenvalues are real), the system exhibits dynamically stable Rabi-like oscillations. Similar to the case in the EP3-based sensor, we choose the time $\chi t=2q\pi$ as the working point.
At $\chi t=2q\pi$, the system evolves back to its initial state.
The noise is then at the vacuum noise level $\mathcal{N}^{2} = (n-1)/2$.
Considering the performance of both signal and noise, we therefore achieve the optimal sensitivity $\delta\epsilon_{\text{opt}}$ of the system at the working point $\chi t=2q\pi$:
\begin{equation}
    \delta\epsilon_{\text{opt}} = \frac{\sqrt{(n-1)/2}}{\mathcal{S}} \propto \epsilon^{2-\frac{1}{n}}.
\end{equation}

In the main text, we provide the scaling of the sensitivity with respect to $\chi$, where $\chi$ determines the time required for the system to evolve back to its initial state.
To better align the analysis of the the sensitivity presented above with the results in main text, we next discuss the connection between $\chi$ and $\epsilon$ in the neighborhood of the EPs.
When the system is sufficiently close to the EPs, the difference between the eigenvalues $\Delta\lambda$ is primarily contributed by the perturbation $\epsilon$ and proportional to $\epsilon^{1/n}$. 
Since the propagation coefficients of the system are composed of a summation of $n$ distinct frequency components, the system's Rabi-like oscillations will undergo collapses and revivals. 
At $t=0$, the system is in its initial state. Subsequently, the magnon modes and the optical field in the cavity undergo parametric oscillations under the SU(1,1)- and SU(2)-type interactions, with the oscillation period determined by $\lambda_0$. However, due to the $n$-fold splitting of eigenvalues caused by $\epsilon$, the system cannot return to its initial state within the period of $\lambda_{0}^{-1}$. Conversely, the system requires multiple oscillation periods of $\lambda_{0}^{-1}$ to return to its initial state. Initially, due to the incoherence between different eigenvalue components, the amplitude of oscillation gradually decreases, preventing the system from returning to its initial state. As time accumulates, coherence is re-established, causing the amplitude to increase again and making the system return to its initial state.
The period of this cycle of collapse and revival is determined by $(\Delta\lambda)^{-1}\sim\epsilon^{-1/n}$, which is:
\begin{equation}
    t = \frac{2q\pi}{\chi} \sim \Delta\lambda^{-1} \sim \epsilon^{-\frac{1}{n}}.
\end{equation}
Therefore, we have $\chi \sim \epsilon^{1/n}$ and obtain the scaling of the sensitivity with respect to $\chi$:
\begin{equation}
    \delta\epsilon_{\text{opt}} \propto \chi^{2n-1},
\end{equation}
which agrees with the results of EP2 \cite{luo2022quantum} and EP3 cases.

\end{widetext}

\end{document}